\documentclass[preprint,review,11pt,authoryear]{elsarticle}
\usepackage[margin=1in]{geometry}

\usepackage[utf8]{inputenc}
\usepackage[version=4]{mhchem}
\usepackage{stackengine}
\usepackage{graphicx}
\usepackage[export]{adjustbox}
\usepackage{float}
\usepackage{array}
\usepackage{xcolor}
\usepackage{color}
\usepackage{tikz}
\usepackage{lineno}
\usepackage{natbib}
\usepackage{nomencl}
\usepackage{hyperref}
\makenomenclature

\usepackage[xindy]{glossaries} 
\newglossaryentry{mesoscopic}{%
  name={Mesoscopic},%
  description={scale refers to finite size description of a system where we account for stochastic factors due to finiteness of the system.}}
  
\newglossaryentry{coarse-grained}{%
  name={Coarse-grained},%
  description={behavior represents average behavior of a group of individuals. For example average direction, velocity, speed etc. These quantities are called coarse-variable or coarse observables or even macroscopic variables.}}
  
\newglossaryentry{microscopic}{%
  name={Microscopic},%
  description={scale behaviors or rules refer to the decisions at individual level in the context of collective behavior. For example following nearby individuals.}}  
  
\newglossaryentry{macroscopic}{%
  name={Macroscopic},%
  description={scale refers to the description of infinite size system that is spatially homogeneous.}}
  
\newglossaryentry{self propelled particles}{%
  name={Self propelled particles},%
  description={are autonomous agents that use energy from the environment into persistent motion.}} 
  
\newglossaryentry{agent based models}{%
  name={Agent based models},%
  description={are computer models of interacting individuals (mobile/non-mobile). }} 
  
\newglossaryentry{hydrodynamic equations}{%
  name={Hydrodynamic equations},%
  description={are mathematical formalism to describe fluid (liquid in motion) dynamics. In collective behavior, a hydrodynamic description considers a mobile animal group as liquid in motion.}} 
  
\newglossaryentry{Master equation}{%
  name={Master equation},%
  description={is an equation describes the temporal evolution of the system in terms of probability of the system being in a given state at a given time ($t$). This probability is given by the summation of different probabilities at a previous time instant ($t'$). The first term is of course that probability that the system was in the given state. Probability of all possible combinations by which the system can transition into the given state in time interval ($t-t'$) are added as second term. Similarly, probabilities of all possible combinations by which the system can transition from given state into other states in time ($t-t'$) are subtracted as third term in the equation.}}
  
\newglossaryentry{Fokker-Planck equations}{%
  name={Fokker-Planck equation},%
  description={is a partial differential equation describing the temporal evolution of probability of a coarse observable that is a stochastic variable. The equation is obtained by retaining the first two terms of the Taylor expansion of the master equation that contains $n^{th}$ order partial derivatives of $n^{th}$ moment of the distribution w.r.t. the coarse observable. Therefore the Fokker-Planck equation contains first and second order partial derivatives of first and second moment of the distribution, respectively.}}  
  
\newglossaryentry{Langevin equation}{%
  name={Langevin equation},%
  description={is a stochastic differential equation describing the temporal evolution of a coarse-variable. The first term is the deterministic term and is often referred to a the drift term. The second term in the equation is the stochastic term and is often called the diffusion term.}}
  
\newglossaryentry{chemical Fokker-Planck Equation}{%
  name={chemical Fokker-Planck Equation},%
  description={is a direct consequence of the chemical Langevin equation.}}
  
\newglossaryentry{chemical Langevin Equation}{%
  name={chemical Langevin Equation},%
  description={scale refers to the description of infinite size system that is spatially homogeneous}}
  
 \newglossaryentry{Order Parameter}{%
  name={Order Parameter},%
  description={scale refers to the description of infinite size system that is spatially homogeneous}}

\makeglossaries

\title{Deriving mesoscopic models of collective behaviour for finite populations}

\author{Jitesh Jhawar$^1$}
\author{Richard G. Morris$^2$}
\author{Vishwesha Guttal$^1$}
\address{1. Centre for Ecological Sciences, Indian Institute of Science, Bengaluru, 560012\\
2. National Centre for Biological Sciences, TIFR, Bengaluru, 560012}

\begin{document}
\begin{abstract}
Animal groups exhibit many emergent properties that are a consequence of local interactions. Linking individual-level behaviour, which is often stochastic and local, to coarse-grained descriptions of animal groups has been a question of fundamental interest from both biological and mathematical perspectives.  In this book chapter, we present two complementary approaches to derive \emph{coarse-grained} descriptions of collective behaviour at so-called \emph{mesoscopic} scales, which account for the stochasticity arising from the finite sizes of animal groups. We construct stochastic differential equations (SDEs) for a coarse-grained variable that describes the order/consensus within a group. The first method of construction is based on van Kampen's system-size expansion of transition rates. The second method employs Gillespie's chemical Langevin equations. We apply these two methods to two \emph{microscopic} models from the literature, in which organisms stochastically interact and choose between two directions/choices of foraging. These `binary-choice' models differ only in the types of interactions between individuals, with one assuming simple pairwise interactions, and the other incorporating ternary effects. In both cases, the derived mesoscopic SDEs have multiplicative/state-dependent noise, {\it i.e.}, the strength of the noise depends on the current state of the system. However, the different models demonstrate the contrasting effects of noise: \emph{increasing} the order/consensus in the pairwise interaction model, whilst \emph{reducing} the order/consensus in the higher-order interaction model. We verify the validity of such mesoscopic behaviour by numerical simulations of the underlying microscopic models.  Although both methods yield identical SDEs for {\it binary}-choice systems that are effectively one-dimensional, the relative tractability of the chemical Langevin approach is beneficial in generalizations to higher-dimensions. We hope that this book chapter provides a pedagogical review of two complementary methods to construct mesoscopic descriptions from microscopic rules, how the noise in mesoscopic models is often multiplicative/state-dependent, and finally, how such noise can have counter-intuitive effects on shaping collective behaviour.   
\end{abstract}

\maketitle

{\it keywords: demographic noise, multiplicative noise, nonlinear dynamics, Fokker-Planck equation, Langevin equation, noise-induced transitions, ecology, collective motion, population dynamics, collective decision making}

\section{Introduction}
Collective behaviour is widespread in the animal kingdom, where many properties of animal groups are emergent, {\it i.e.}, they often arise from simple local interactions among individuals~\citep{vicsek1995,couzin2002,parrish2002self,sumpter2010principles,sumpter2006philtrans}. These emergent properties include, for example, visually fantastic patterns of flocks of starlings swirling in synchrony~\citep{cavagna2010pnas,attanasi2014}, locusts on the march~\citep{uvarov1977,simpson2008currbiol,buhl2006}, conflict resolution and consensus decision making in groups of animals~\citep{deneubourg1989tree,petit2010,beckers1990collective,pratt2002bes,couzin2005nature}, collective navigation~\citep{grunbaum1998evoeco,guttal2010,berdahl2013}, or colonies of social insects performing complex tasks~\citep{seeley1989honey,franks1992self,dussutour2006bes,schultheiss2012,gordon1996nature,bonabeau1997tree}.

Understanding these emergent properties has proved to be a topic of great interest in both biology and the physical sciences~\citep{flierl1999individuals,camazine2003self,sumpter2010principles,guttal2014ecology}. Broadly speaking, the focus of biologists has typically been to understand how organisms interact; processing local information from their neighbours. This information is often noisy and inaccurate, yet repeated interactions amongst individuals can result in the aforementioned collective properties. Physicists and mathematicians, by contrast, have tried to develop simple theories to describe how local interactions between such agents can combine to result in larger scale properties of groups and populations. These two approaches are almost complementary to each other, with theory able to make connections between the observed meso/macroscopic behaviours and the likely individual-level decisions of organisms; the latter being difficult to measure directly in the natural setting of a group.

Here, we will use the word microscopic to refer to behaviour at the individual level ($N=1$), and the word macroscopic to describe the behvaiour of infinite sized systems $(N \to \infty)$. The mesoscopic scale is in between these two, and corresponds to intermediate (large but finite) group sizes.  The aim of this book chapter is to introduce readers to a couple of complementary analytical methods that help us link individual behaviour to emergent properties of the collective. In doing so, our focus will be on the mesoscopic scale: aiming to understand the important role of stochasticity resulting from finite group sizes.
\\ 
\section{Background}

For historical context, we begin by briefly presenting the classic {\it Vicsek model} of collective motion~\citep{vicsek1995}, which has had a profound impact on the field of collective behavior. It has inspired a large body of theoretical and empirical research in physics, mathematics and biology. In this simple model, we consider a population of $N$ self-propelled particles (SPP) moving with constant speed $s$ in a continuous space of volume $L^d$, where $d$ is the spatial dimension. (For the purpose of this chapter we restrict ourselves to one- and two-dimensions only). Boundary conditions are assumed to be periodic, implying that a particle leaving the boundary at the right-extreme returns to the left-extreme edge of the space, \emph{etc}. Topologically, a one-dimensional system with periodic boundaries is equivalent to a ring whereas a two-dimensional system is equivalent to a torus. 

The basic rule of the model is that each individual, $i$, moves in the average direction of its {\it neighbors}, plus some small error.  More specifically, at any given time $t$, the directions of all individuals $\left\{\theta_i:\, i=1,\ldots,N\right\}$ are updated synchronously according to 
\begin{equation}
\label{eq:vicsek}
\theta_i(t+\delta t) = \overline{\theta (t)} + \Delta \theta (t),
\end{equation}
\noindent where the overbar represents an average over individuals within a radius $r$ from the location of individual $i$, the so-called {\it focal} individual. The first term on the right-hand side captures how the focal individual responds to interactions with its neighbours, and corresponds to calculating $\arctan\left( \overline{\sin\theta}  / \overline{\cos\theta}\right)$. The second term is a noise term representing the error a focal individual makes while copying the average neighborhood direction.  $\Delta \theta(t)$ is therefore a random variable, typically chosen from a uniform distribution over the range $[-\eta/2,\eta/2]$, where $\eta\le 2\pi$. Once the directions $\theta_{i}(t+\delta t)$ have been calculated, each individual moves in that direction with speed $s$ for a small time interval $\delta t$, after which all angles are re-computed. 

\begin{figure}[!]
\begin{adjustbox}{addcode={\begin{minipage}{\width}}{\caption{Various patterns of collective motion produced by the classic Vicsek model of collective motion. Reproduced here with permission from {\it American Physical Society: Physical Review Letters}~\citep{vicsek1995}. There are 300 particles in each of these panels, with the arrow indicating the current direction of motion and the short solid lines preceding it indicates the trajectory of that particle over last 20 time steps. (a) High density and large-noise ($L=7$, $\eta=2.0$), but initial stages of simulation. (b) Low-density and low-noise ($L=25$, $\eta=0.01$) lead to formation of groups that move in random directions. (c) High-density and large-noise (same as parameters in (a): $L=7$, $\eta=2.0$), but after some time and (d) High-density and low-noise ($L=5$, $\eta=0.1$) results in highly ordered motion.
\label{fig:vicsek-patterns}}
\end{minipage}},rotate=0,center}
\includegraphics[scale = 0.65]{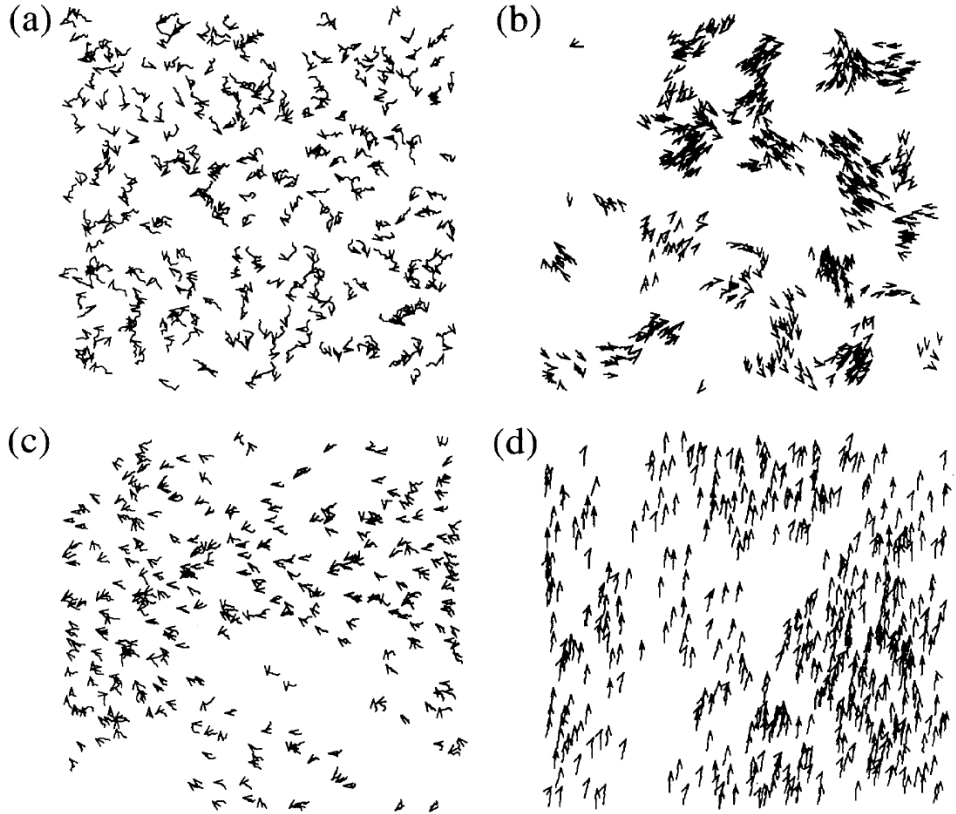}
\end{adjustbox}
\end{figure}

\begin{figure}[!]
\begin{adjustbox}{addcode={\begin{minipage}{\width}}{\caption{Phase diagram of the Vicsek model, reproduced here with permission from{\it American Physical Society: Physical Review Letters}~\citep{vicsek1995}. The order parameter (denoted as $v_a$ in the Figure above; we adopt the notation $m$ in this manuscript for the equivalent quantity) which captures the degree of alignment or polarisation of the entire population of particles. It shows features of phase-transition from order to disorder as the strength of noise ($\eta$) increases in (a) and as the density of particles ($\rho$) increases in (b). See~\cite{vicsek1995} for parameter values and other details. 
\label{fig:vicsek-phasedia}}
\end{minipage}},rotate=0,center}
\includegraphics[scale = 0.55]{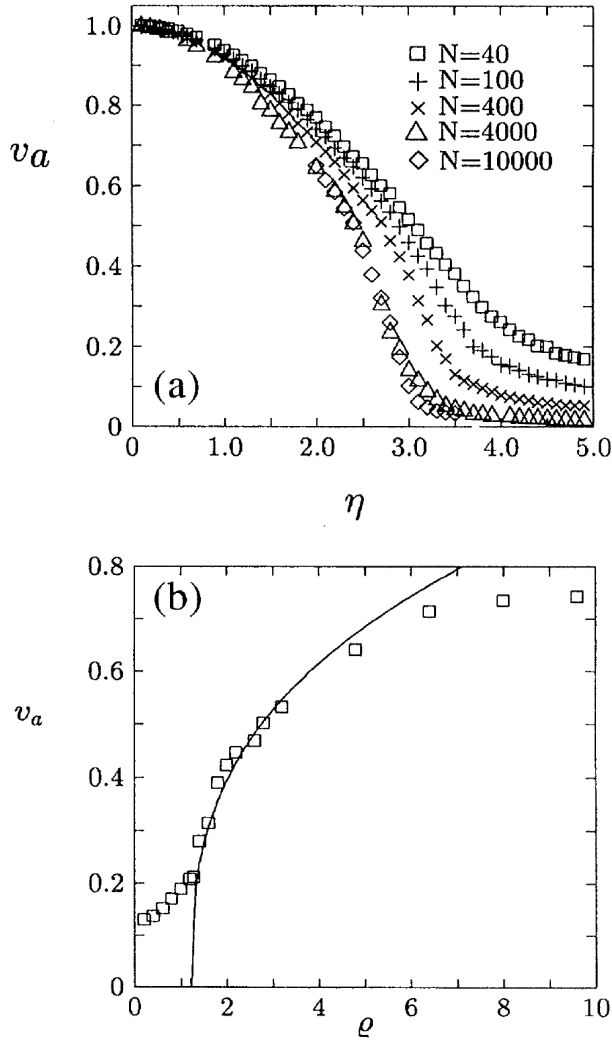}
\end{adjustbox}
\end{figure}

The Viscek model is therefore a simple model of SPPs in continuous space, where each individual updates their direction of motion synchronously at discrete time steps based on Eq.~\eqref{eq:vicsek}. Nevertheless, the model exhibits a surprisingly rich variety of spatial patterns of collective movement. For example, it is evident from Fig.~\ref{fig:vicsek-patterns} that ordering is facilitated by both low levels of noise and high densities of individuals. By calculating the average alignment of the individuals' motion, Viscek defined a global `order parameter' in analogy with the polarization of a ferromagnet.  Mathematically, this is given by the scalar quantity
\begin{equation}\label{eq:vicsek-op}
m(t) = \frac{1}{N\,s}\bigg\vert \sum_{i=1}^N \mathbf{v_i}(t) \bigg\vert, 
\end{equation}
where $\mathbf{v_i}$ represents the velocity vector of individual $i$ at time $t$. The normalisation includes $s$, which is the speed of each individual. The quantity $m$ is popularly known as the Vicsek order-parameter in the physics literature; when close to $0$ the group exhibits disordered motion, whilst when close to $1$ the group is highly polarised, {\it i.e.}, all individuals move in the same direction. Notably, \cite{vicsek1995} showed via numerical simulations that the ensemble average of this quantity exhibits a phase transition in two-dimensions as a function of increasing density, and decreasing noise (Fig.~\ref{fig:vicsek-phasedia}).

Since Vicsek's original paper, an array of similar SPP models of collective motion have been developed in continuous and discrete one-, two- and three-dimensional space~\citep{mogilner1999non,czirok1999collective,inada2002jtb,chowdhury2002jphysics,couzin2002,gregoire2004,nishinari2006modelling,chate2008,strombom2011collective}. Several of these studies rely on numerical simulations to elucidate how simple microscopic interactions result in large scale collective motion. A number of analytical theories have been developed to describe the behaviour of the order-parameter $m$ in Vicsek-like and other collective motion models~\citep{toner1995long,czirok19991D,jadbabaie2003coordination,ginelli2016physics}. However, much of their focus is on understanding behaviour at the macroscopic scale {\it i.e.}, in the limit of infinite population/group sizes, ignoring the role of stochastic effects arising from finite population/group sizes~\citep{kunwar2004collective,aldana2007phase,baglietto2009nature,ramaswamy2010annrev,romanczuk2012mean,marchetti2013rmp}.


Here, our aim is to demonstrate a `first principles' approach to the derivation of \emph{mesoscopic} models for the order parameter $m$, where noise is a function of both individual behaviours (which are stochastic) and finite size of groups. This is in contrast to the many {\it ad-hoc} methods of including stochastic effects, such as simply appending a noise term to otherwise mean-field dynamics, for example.  For analytical tractability, we will demonstrate the construction of such mesoscopic descriptions from microscopic rules that are simpler than the Vicsek model. Our aim is that this chapter be accessible to readers at the graduate, or even advanced undergraduate, level in numerate subjects like mathematics, physics, statistics and quantitative biology. Although we endeavour to explain most concepts and techniques as they are introduced (occasionally referring to standard books for certain technical/detailed steps), basic familiarity of the reader with the following subjects will be helpful: dynamical-system based analyses of ordinary differential equations (ODEs) \citep{strogatz2018}, probability theory, stochastic differential equations (SDEs), Ito-calculus, Master equations and Fokker-Planck equations~\citep{gardiner2009}.

Before proceeding further, we state our goals in more specific terms. Given a microscopic model of interactions between individuals, we would like to derive an SDE for a coarse variable that quantifies some aspect of the collective. If the coarse-variable is $m$, at mesoscopic scales where stochasticity is important, we expect the dynamical equation to be of the form
\begin{equation}
\frac{dm}{dt} = f(m) + g(m)\, \eta(t),
\label{eq:mdot}
\end{equation}
where $\eta(t)$ is so-called delta-correlated Gaussian white noise with zero mean--- {\it i.e.}, $\langle \eta(t) \rangle = 0$ and $\langle \eta(t)\,\eta(t^\prime) \rangle = \delta(t-t^\prime)$ [see~\cite{gardiner2009} for more details].  This is an example of an SDE, and is sometimes also referred to as a \gls{Langevin equation}.  The function $f(m)$ is referred to as the {\it determinsitic} or {\it drift} term, whilst $g(m)$ is often called the {\it stochastic} or {\it diffusion} term~\citep{kolpas2007pnas,yates2009pnas,kolpas2008thesis}. If $g(m)$ is constant, the strength of noise is independent of the state of the system, and noise can be thought of as external, or \emph{additive}, to the deterministic dynamics. On the other hand, if $g(m)$ is not constant, the strength of the noise depends on the current state $m(t)$. Such noise is called multiplicative, or state-dependent, and is known to produce unexpected results~\citep{horsthemke1984,altschuler2008spontaneous,lawson2013spatial,boettiger2018noise}.  In the following sections, we will not only describe how to derive equations of the form (\ref{eq:mdot}) from given probabilistic rules for individual-level behaviour, but also provide some heuristic insight into the different types of behaviours that can arise.

\section{Mesoscopic description of a pairwise binary-choice model}
\label{sec:1d-pairwise}
In this section, we illustrate two complementary methods for deriving mesoscopic dynamical descriptions of collective behaviour from first principles. To do so, we choose a simple microscopic model of binary-choice, used to describe foraging animals that must choose between two food sources~\citep{kirman1993quartecon,biancalani2014prl}, but also applied in several other diverse contexts, such as traders in financial markets~\citep{kirman1993quartecon,alfarano2008econdyn}, and the recruitment of cell signaling molecules~\citep{altschuler2008spontaneous}. Here, individuals can collect food from one of two sources, labeled by $i=1,2$, which can be thought of as being located at either end of a long line. The food source to which each individual is moving therefore determines its direction, either positive or negative, which is denoted by $X_i$. The proportion of the colony (total number of individuals, $N$) moving in a given direction is written using lowercase $x_i=N_i/N$, where $N_i$ is simply the number of individuals with direction of motion $X_i$.

\subsection{The model}\label{sec:pairwise-antmodel-1d}
As with any model of collective decision making, the choice that any individual makes is influenced by its {\it interactions} with the group ({\it i.e.}, what others are doing) as well as external factors.  In this section, we focus first on pairwise interactions, with higher-order interactions (specifically, ternary interactions) described in Section~\ref{sec:higher-order-locustmodel-1d}.  For the case of our pairwise binary-choice model, we imagine a continuous-time stochastic protocol whereby individuals can interact in two different ways.

Firstly, individuals can {\it copy} each other.  For example, an interaction between a focal individual moving towards source $1$ with another moving towards source $2$ causes the focal individual to copy the other, and therefore switch to move towards the other source, $2$. Likewise, an interaction between a focal individual of direction $X_2$ with another of direction $X_1$ causes the focal individual to switch to move towards source $1$. These interactions can be represented in the same way as chemical reactions, writing:
\begin{subequations}\label{eq:copying}
\begin{equation}
X_1 + X_2 \xrightarrow {c} 2 X_2, 
\label{eq:copyingA}
\end{equation}
and
\begin{equation}
X_2 + X_1 \xrightarrow {c} 2 X_1.
\label{eq:copyingB}
\end{equation}
\end{subequations}
\noindent Such copying interactions, occurring at a specific rate $c$, lead to an auto-catalytic process wherein an individual's chance of switching directions depends on the proportion of individuals already moving in that direction.

Secondly, the model also incorporates a level of {\it random} choice, where individuals ignore the actions of the others, and switch from their current food source spontaneously at a rate $s$. The corresponding `chemical' reactions are:
\begin{subequations}\label{eq:stochastic}
\begin{equation}
X_1 \xrightarrow {s} X_2, 
\label{eq:stochasticA}
\end{equation}
and
\begin{equation}
X_2 \xrightarrow {s} X_1.
\label{eq:stochasticB}
\end{equation}
\end{subequations}

Since the proportions of individuals going towards either source sum to unity, {\it i.e.}, $x_1 + x_2 = 1$, we can define a {\it coarse-variable} $m = x_1 - x_2 = 1-2x_1$, which takes values from $-1$ to $1$ and encapsulates the propensity of the group to be moving to either of the food sources. The extreme values of $m$ correspond to consensus decision making, {\it i.e.}, all individuals collectively forage from source $1$ ($m=1$) or source $2$ ($m=-1$). The values of $m$ close to zero imply a lack of consensus, or collective behaviour, with equal proportions going towards either of the food sources. In-line with the previous section, we refer to $m$ as the order parameter.

Indeed, let us compare the pairwise interaction binary-choice model (sometimes referred to as the `ant model' in the literature) with the Vicsek model. The pairwise interaction model is a highly simplified representation of collective behaviour, analogous to both the classical Voter model~\citep{cox1986diffusive} and the stepping stone model of population genetics~\citep{kimura1964stepping}. Unlike the Vicsek model, the rules are described via probabilistic reactions, rather than a difference equation. Further, here we ignore space, and assume that any individual can interact with any other individual in the population. Additionally, the update protocol is asynchronous {\it i.e.}, individuals do not update their orientations simultaneously; this is in contrast to the synchronous update of orientations and positions of all individuals in the Vicsek model. Further, whilst the Vicsek model considers a continuous $\theta$ (the angle/direction of movement), this model assumes only two sources of food, and hence two directions. (For completeness, we present an extension to the pairwise interaction model in~\ref{app:2d-pairwise}, where we permit four choices, in order to model movement in a two-dimensional space). We also note that the order parameter of the Vicsek model ranges from 0 (disorder) to 1 (order), whereas here, it ranges from $-1$ to $1$, with both extremes representing order (consensus) and the value $m=0$ representing disorder (or lack of consensus). 

Taken together, the aforementioned simplifications make the task of formally constructing mesoscopic equations relatively straightforward, whereas coarse-grained descriptions of the Vicsek model typically rely on symmetry arguments~\citep{toner1995long,ramaswamy2010annrev} and are harder to explicitly derive from the underlying microscopic model.

\subsection{Constructing mesoscopic SDEs}
\label{sec:coarse-grained-equations-pairwise}
We describe two complementary approaches to constructing a mesoscopic dynamical equation for the coarse-grained variable $m$, of the form (\ref{eq:mdot}). The first method is based on van Kampen's system-size expansion method \citep{van1992stochastic}, and is well known in the statistical physics community.  Amongst other groups, the technique has been adopted extensively by McKane and colleagues for use in a variety of biological, ecological and epidemiological contexts \citep{mckane2004,mckane2005prl,biancalani2014prl}. The second approach is called the chemical Langevin equations (CLEs), developed by \cite{gillespie2000}. The latter method is prevalent in chemistry \citep{sotiropoulos2011analytical} and to some extent in the biological literature concerning gene regulatory networks~\citep{simpson2009noise,el2006advanced}, biochemical pathway analysis~\citep{rudiger2014stochastic} and epidemiology~\citep{colizza2007modeling,yuan2011stochastic,wang2016statistical}. However, aside from a couple of studies~\citep{datta2010,joshi2018Demographic} it is, to our knowledge, rarely used in the ecology literature.

\subsubsection{van Kampen's system-size expansion of transition rates}
\label{sec:sys-size-1D-pairwise}
We begin by summarising the step-by-step approach of this method: given a microscopic model, or protocol, the first step is to write how individual/microscopic rules translate into \emph{transition rates} between different states of the system [denoted by the shorthand $\boldsymbol{x}=\{x_1,x_2\}^\mathsf{T}$]. The transition rates depend on the current state of the system and the system size (in this case, the number of individuals, $N$). In the second step, using these transition rates, we write a \emph{master equation} for the temporal evolution of the probability density function (PDF) for the system's state, $P(\boldsymbol{x},t)$. In the third step, we write the transition rates in terms of operators and Taylor-expand them for large but finite $N$, retaining only the two leading order terms. This provides a so-called \emph{Fokker-Planck} equation for $P(\boldsymbol{x},t)$, which approximates the aforementioned master equation. Then, we transform the state variables $\boldsymbol{x}$ to the coarse-variable of interest (in this case, $m$). Finally, we construct the SDE, or Langevin equation, that corresponds to the transformed Fokker-Planck equation.

{\it Transition rates:} For the model under consideration, since the individuals are indistinguishable, the state of the system is given by the variables $x_1$ and $x_2$: the proportion of individuals moving in the negative and positive directions, respectively. Both $x_1$ and $x_2$ take discrete values in the range $\{0,1/N,\ldots,N-1/N,1\}$.  Given an initial state $(x_1, x_2)$, a single reaction of the type (\ref{eq:copying}a) results in the state $(x_1-1/N, x_2+1/N)$, whilst (\ref{eq:copying}b) results in $(x_1+1/N, x_2-1/N)$. Both transitions occur at a rate $c\,x_1\,x_2$, {\it i.e.}, the probability of picking two random individuals moving in different directions, multiplied by the rate at which they copy each other.  Similarly, a single reaction of type (\ref{eq:stochastic}a) also results in the state $(x_1-1/N, x_2+1/N)$, whilst (\ref{eq:stochastic}b) results in $(x_1+1/N, x_2-1/N)$.  Here the transitions occur at a rate $s\,x_1$ and $s\,x_2$, respectively, which is just the probability of picking a given individual, multiplied by the rate at which they switch direction. Combining the above, we write down the overall rate associated with a transition from state $(x_1,x_2)$ to $(x_1-1/N, x_2+1/N)$, which is just
\begin{equation}
T_{21}=T\left(x_1-\frac{1}{N}, x_2+\frac{1}{N} \,\bigg\vert \, x_1, x_2\right) = c\, x_1\, x_2 + s\, x_1.\label{eq:T12}
\end{equation}
Similarly, for the transition $(x_1,x_2)$ to $(x_1+1/N, x_2-1/N)$, we have
\begin{equation}
T_{12}=T\left(x_1+\frac{1}{N}, x_2-\frac{1}{N} \,\bigg\vert \, x_1, x_2\right)  = c\, x_2\, x_1 + s\, x_2.\label{eq:T21}
\end{equation}
Here, the shorthand notation $T_{ij}=T(x_i+1/N, x_j-1/N\, \vert \, x_i,x_j)$ has been introduced for convenience in the coming manipulations.  We also note that, due to the constraint $x_1+x_2=1$, technically only one of $x_1$ or $x_2$ is required to fully determine the state of the system, a point to which we will return later.

{\it Master equation:} We describe the evolution, or rate of change, of $P(\boldsymbol{x}, t)$ by the influx of density associated with the state $\boldsymbol{x}$ from that of the other states $\boldsymbol{x}^\prime\neq\boldsymbol{x}$, minus the outflow of density from state $\boldsymbol{x}$ to the other states $\boldsymbol{x}^\prime\neq\boldsymbol{x}$. The resulting partial differential equation is called a master equation and is given by
\begin{eqnarray}\label{eq:master-eq}
\frac{\partial P(\boldsymbol{x}, t)}{\partial t} & = & \sum_{\boldsymbol{x}^\prime\neq\boldsymbol{x}}\left[T(\boldsymbol{x}\,\vert\,\boldsymbol{x}^\prime)\,P(\boldsymbol{x}^\prime,t) - T(\boldsymbol{x}^\prime\,\vert\,\boldsymbol{x})\,P(\boldsymbol{x},t)\right]\\ 
& = & \sum_{x_1' \ne x_1} \sum_{x_2' \ne x_2} 
\left[T\left(x_1, x_2 \, \vert\, x_1', x_2'\right)\,  P(x_1', x_2',t) -  
T\left( x_1', x_2' \, \vert \, x_1, x_2\right) \, P(x_1, x_2,t)\right] \\
& = & \sum_{x_1' \ne x_1} \sum_{x_2' \ne x_2} 
\left[T\left(x_1, x_2\, \vert \, x_1', x_2'\right)  P(x_1', x_2',t)\right] - \left(T_{12}+T_{21}\right) P(x_1, x_2,t). \label{eq:simplified-master}
\label{eq:masterEquation}
\end{eqnarray}
{\it Step operators:} The next step is to rewrite the master equation in terms of step-operators, $\mathcal{E}_i^{\pm}$, which are defined as
\begin{equation}\label{eq:stepop}
\mathcal{E}_i^\pm f(x_i) = f(x_i \pm \frac{1}{N}).
\end{equation} 
The key idea is to use such step operators to compactly replace the sum that appears in Eq.~\eqref{eq:simplified-master}. For example, the first term of Eq.~\eqref{eq:simplified-master} can be drastically simplified by using both the aforementioned step operators and the previously introduced shorthand for transition rates:
\begin{eqnarray}\label{eq:inflow-operators}
\sum_{x_1' \ne x_1}\sum_{x_2' \ne x_2} T\left(x_1, x_2\,\vert \, x_1', x_2'\right) \, P(x_1', x_2',t) 
& = & T\left(x_1, x_2\,\bigg\vert\, x_1-\frac{1}{N}, x_2+\frac{1}{N}\right) \, P(x_1-\frac{1}{N}, x_2+\frac{1}{N},t) \nonumber \\ 
& \quad & + T\left(x_1, x_2\,\bigg\vert\, x_1+\frac{1}{N}, x_2-\frac{1}{N}\right)\, P(x_1+\frac{1}{N}, x_2-\frac{1}{N},t) \nonumber\\
& = & \mathcal{E}_1^- \mathcal{E}_2^+\, T_{12}(x_1,x_2)\, P (x_1,x_2,t) \nonumber\\
&\quad& +\mathcal{E}_1^+ \mathcal{E}_2^- \,T_{21}(x_1,x_2)\, P (x_1,x_2,t).
\end{eqnarray}
\noindent Note that the operators act on all functions, and all arguments therein, that appear to its right hand side.  The full master equation, now in the form of operators, is given by
\begin{equation}\label{eq:master-op}
\frac{\partial P(x_1, x_2, t)}{\partial t} = (\mathcal{E}_1^- \mathcal{E}_2^+ - 1) T_{12} P (x_1,x_2,t) + 
(\mathcal{E}_1^+ \mathcal{E}_2^- - 1)T_{21} P (x_1,x_2,t), 
\end{equation} 
where the functional dependence of the rates $T_{12}$ and $T_{21}$ on the state $(x_1, x_2)$ have been dropped for brevity.

{\it Taylor expansion:} We now recognise that for large $N$ the action of the step operators may be approximated by a Taylor series.  In doing so, we implicitly consider a continuous approximation of the variables $x_i$.  Retaining only the first two terms in the expansion, we use the generic form
\begin{equation}\label{eq:step_op}
\mathcal{E}_i^\pm f(x_i) = f(x_i \pm \frac{1}{N}) = 
\left[1 \pm \frac{1}{N} \frac{\partial}{\partial x_i} + \frac{1}{2N^2} \frac{\partial^2}{\partial x_i^2}\right]\,f(x_i). 
\end{equation}
Applying this expansion scheme to Eq.~\eqref{eq:master-op} and simplifying, we get
\begin{equation}
\begin{split}
\frac{\partial P(x_1,x_2,t)}{\partial t} = & \; \frac{1}{N} \left(\frac{\partial}{\partial x_2} - \frac{\partial}{\partial x_1}\right)[T_{12}P(x_1,x_2,t)] + \left(\frac{\partial}{\partial x_1} - \frac{\partial}{\partial x_2}\right)[T_{21}P(x_1,x_2,t)] \,+ \\ 
& \frac{1}{2N^2}\left(\frac{\partial}{\partial x_1} - \frac{\partial}{\partial x_2}\right)^2[(T_{12} + T_{21})P(x_1,x_2,t)].
\end{split}
\label{eq:fokker-Planck-transition-rates12}
\end{equation}
Rearranging various terms, rescaling time according to $\; t' = t/N$ and then dropping the prime for ease of notation, we get the canonical form of a Fokker-Planck equation
\begin{equation}
\begin{split}
\frac{\partial P(x_1,x_2,t)}{\partial t} = & \left[ - \sum_{i=1}^2\frac{\partial [\mathcal{A}_i\:P(x_1,x_2,t)]}{\partial x_i} + \frac{1}{2N}\sum_{i,j=1}^2\frac{\partial^2\: [\mathcal{B}_{ij}\:P(x_1,x_2,t)]}{\partial x_ix_j}\right],
\end{split}
\label{eq:fokkerPlanck-canonical-1d}
\end{equation}
where $\mathcal{A}$ is vector and $\mathcal{B}$ is a matrix whose elements are given, respectively, by
\begin{equation}\label{eq:drift-diffn-1d}
\mathcal{A}_i  =  T_{ij}-T_{ji},\ \mathrm{and}\ \mathcal{B}_{ij}  =  (T_{ij}+T_{ji})(-1)^{i+j}.
\end{equation}
\noindent Notice that our derivation of the Fokker-Planck equation~\eqref{eq:fokkerPlanck-canonical-1d} and the associated terms in Eqs.~\eqref{eq:drift-diffn-1d} make little reference to the pairwise interaction model. In other words, the above equations describing the time evolution of the PDF are based on very general features, and largely independent of the specific model under consideration. As a result, we may first apply this to the pairwise binary-choice model, before returning to re-use the same expression in Section~\ref{sec:higher-order-locustmodel-1d} for the extension to ternary interactions.  For the pairwise model, substituting Eqs.~\eqref{eq:T12} and~\eqref{eq:T21} into Eqs.~\eqref{eq:drift-diffn-1d}, we get
\begin{equation}
\mathcal{A}_1  =  s(x_2-x_1) = -\mathcal{A}_2,\label{eq:ant-model-drift-difn-1d-AB}\ \mathrm{and}\ 
\mathcal{B}_{ij}  =  \left[2cx_1x_2 + s(x_1+x_2)\right] (-1)^{i+j}.
\end{equation}

\textit{Mesoscopic SDE}: Given a Fokker-Planck equation for a PDF, it is, in principle, always possible to construct a system of corresponding SDEs in order to describe the stochastic variables whose statistics are described by that PDF. For details, we refer the reader to the classic book on stochastic processes by~\cite{gardiner2009}. For the pairwise interaction model, the system of SDEs that are equivalent to a Fokker-Planck equation with the terms \eqref{eq:ant-model-drift-difn-1d-AB}, are:
\begin{equation}\label{eq:equivalentLangevin1D}
\frac{dx_1}{dt}  =  \mathcal{A}_1 + \frac{1}{\sqrt[]{N}}\xi_1(t),\ \mathrm{and}\ \frac{dx_2}{dt}  =  \mathcal{A}_2 + \frac{1}{\sqrt[]{N}}\xi_2(t),\ \mathrm{where}\ \langle\xi_i(t)\xi_j(t')\rangle  =  \mathcal{B}_{ij}\delta(t-t').
\end{equation}
In the above SDEs, $\xi_1$ and $\xi_2$ represent noise sources that are correlated. Moreover, $x_1$ and $x_2$ are constrained by the fact that $x_1+x_2=1$, and it is therefore possible to eliminate one of the two variables. To obtain a simpler SDE, which is still consistent with a Fokker-Planck description involving the coefficients~\eqref{eq:ant-model-drift-difn-1d-AB}, we therefore invoke the transformation $\xi_i = \sum_{j=1}^2 \mathcal{G}_{ij}\eta_j$, where $\mathcal{B} = \mathcal{G}\mathcal{G}^T$ and the $\eta_j$ ($j=1,2$) each correspond to independent delta-correlated Gaussian white noises. Because $\mathcal{B}$ and $\mathcal{G}$ are square $2 \times 2$ matrices, it is easy to see that $\mathcal{G}_{ij} = (-1)^{i+j+1}\sqrt[]{2cx_1x_2+s(x_1+x_2)}/\sqrt[]{2}$. Substituting this into Eqs.~\eqref{eq:equivalentLangevin1D} we obtain, 
\begin{equation}
\frac{dx_1}{dt} = s(x_2-x_1) + \sqrt{\frac{2c x_1x_2+s(x_1+x_2)}{2N}}\left[\eta_2(t)-\eta_1(t)\right] = -\frac{dx_2}{dt}.
\label{eq:Langevin1D}
\end{equation}
Introducing $m=x_1-x_2$ then yields
\begin{equation}
\frac{dm}{dt} = -2sm + \frac{1}{\sqrt[]{N}}\,\sqrt[]{c(1-m^2)+2s}\,\left[\eta_2(t)-\eta_1(t)\right].
\end{equation}
\noindent We use the fact that the difference between two variables, each drawn from independent but identical Gaussian distributions, is statistically equivalent to simply drawing from a single Gaussian distribution. Therefore, the above SDE simplifies to 
\begin{eqnarray}
\frac{dm}{dt} = -2sm + \sqrt[]{\frac{2}{N}}\,\sqrt[]{c(1-m^2)+2s}\,\eta(t).\label{eq:mdotNoRescale}
\end{eqnarray}
Finally, it is helpful to move to a rescaled time $t' = 2st$.  However, since the variance of the noise term depends on the timescale, care must be taken.  Without wanting to digress too far on the subject of SDEs, the easiest course of action is to rescale the Fokker-Planck equation corresponding to \eqref{eq:mdotNoRescale} by multiplying it with $dt/dt^\prime = 1/2s$.  Relabeling $t'$ as $t$ and writing-down a new SDE, we get 
\begin{equation}
\label{eq:ant-model-langevin-syssize}
\frac{dm}{dt} = -m + \sqrt[]{\frac{N_c}{N}}\,\sqrt[]{1-m^2+2s}\,\eta(t),
\end{equation}
where $s/c=s$ and $N_c=1/s$~\citep{biancalani2014prl}. As before, $\eta(t)$ is delta-correlated Gaussian white noise, {\it i.e.}, $\langle \eta(t) \rangle = 0$ and $\langle \eta(t)\,\eta(t^\prime) \rangle = \delta(t-t^\prime)$. 

In the Section~\ref{sec:solving-coarse-grained-1D}, we continue the analyses of Eq.~\eqref{eq:ant-model-langevin-syssize} by solving for the steady-state solution of the corresponding Fokker-Planck equation and by generating sample stochastic trajectories numerically. Before doing so, however, we demonstrate an alternative, and often complementary approach to the derivation of mesoscopic SDEs, known as the chemical Langevin equation approach~\citep{gillespie2000}.

\subsubsection{Gillespie's chemical Langevin approach} \label{sec:chem-lagenvin-ant-model-1d}

An alternative method for the construction of mesoscopic equations was developed by~\cite{gillespie2000} in the context of many interacting chemical species, which provides a neat heuristic to write down SDEs directly from a set of chemical reactions.  Here, as with van Kampen's approach, the method accounts for the stochasticity that arises from finite population sizes. The mathematical arguments for the approach are described in detail by~\cite{gillespie2000}, where both the similarities and differences to van Kampen's method are discussed at length.  For completeness, we tersely recapitulate some of the arguments/steps in~\ref{app:chemlangevin}. Given certain assumptions are satisfied, the procedure involves constructing so called {\it propensity functions} and {\it state change vectors/matrices}, and then simply assembling the SDE accordingly. We demonstrate this approach in the context of the pairwise interaction model described in Section~\ref{sec:pairwise-antmodel-1d}. 

{\it Propensity functions:} The propensity function, denoted by $a_j(\boldsymbol{x})$, represents the probability per unit time of reaction $j$. Here, as before, $\boldsymbol{x} =\{x_1,x_2\}^\mathsf{T}$, which represents the state of the system in terms of the concentration/proportion of individuals of different species.  [Note that in \ref{app:chemlangevin} we follow \cite{gillespie2000} and define the propensity function in terms of the absolute number of individuals ($N_i$) of each species present in the system, rather than concentrations, which are used here for comparison with van Kampen's approach]. The reaction propensities are computed based on the principle of mass action, {\it i.e.}, that the probability of a reaction occurring at a given instant is equal to a specific rate multiplied by the product of the concentrations of all the constituent chemicals of that reaction at that time. Let us recall that the pairwise interaction model is described by the four microscopic rules or `chemical reactions' set out in Eqs.~(\ref{eq:copyingA}--\ref{eq:stochasticB}). Consider the first reaction, which represents an individual originally moving towards source $1$, who copies the behaviour of another individual and changes to move towards source $2$.  The corresponding propensity function is just the copying rate $c$ multiplied by the product of the proportion of individuals moving in direction $1$ ($x_1$) and the proportion of individuals moving in direction $2$ ($x_2$); thus, $a_1(\boldsymbol{x}) = c\:x_1x_2$. Applying this procedure to the remaining reactions [(\ref{eq:copyingB}), (\ref{eq:stochasticA}) and (\ref{eq:stochasticB})] yields the following propensity functions:
\begin{equation}
\begin{split}
& a_2(\boldsymbol{x}) = c\,x_1x_2, \nonumber \\
& a_3(\boldsymbol{x}) = s\,x_1, \nonumber \\
& a_4(\boldsymbol{x}) = s\,x_2. \nonumber
\end{split}
\end{equation}

{\it State change matrix}: We then construct the state change, or stoichiometry, matrix $\boldsymbol{\nu}$, of which an element $\nu_{ji}$ represents the change in the number of individuals of species $i$ when the reaction $j$ occurs. The rows of $\boldsymbol{\nu}$ therefore correspond to the reactions and the columns represent the different species of the individuals.  For the pairwise interaction model, for example, the first element in the second row represents a loss of one individual moving towards food source 1, and hence $\nu_{21}=-1$.  Conversely, the second element in the second row represents a gain of an individual moving towards source 2, and hence $\nu_{22}=1$. Accounting for all four reactions, $\boldsymbol{\nu}$ is a $4 \times 2$ matrix, given by
\begin{equation}
\boldsymbol{\nu} = \begin{bmatrix}
-1 & 1 \\
1 & -1 \\
-1 & 1 \\
1 & -1
\end{bmatrix}.
\end{equation}
 
{\it Chemical Langevin equations:} Using the propensity functions and the state change matrix, we can follow Gillespie's recipe and write-down the chemical Langevin equation(s). While we refer the reader to the mathematical arguments in~\citep{gillespie2000} and \ref{app:chemlangevin}, the basic idea is as follows. Each reaction $j$ causes a change in the concentration of species $i$ per unit time, which is given by the product of rate of that reaction [{\it i.e.}, the propensity function $a_j(\boldsymbol{x})$] with the associated change in that species numbers ({\it i.e.}, the stoichiometry matrix entry $\nu_{ji}$). Thus, the expected rate of change in the concentration of species $i$ due to reaction $j$ will be $\nu_{ji} a_j(\boldsymbol{x})$. However, chemical reactions are stochastic events and the above term yields only an expected number of reactions per unit time. To take account of stochastic fluctuations, Gillespie uses the fact that, for large enough $N$, the change in the concentration of species $i$ due to reaction $j$ can be approximated by a normally distributed random variable for which the variance and the mean are both equal to the the propensity of the reaction $j$.  Applying this principle to all reactions $j$ that result in a change in concentration of chemical $i$, we obtain the system of `chemical' Langevin equations
\begin{equation}
\frac{d x_i(t)}{dt} = \sum_{j=1}^r \nu_{ji} a_j(\boldsymbol{x})  + \frac{1}{\sqrt[]{N}}\sum_{j=1}^r \nu_{ji}[a_j(\boldsymbol{x}(t))]^{1/2}\mathcal{\eta}_j(t),\ \textrm{for}\, i = 1,...,n.
\end{equation}
where $n$ is the total number of species/chemicals, $r$ is the total number of reactions, $N$ is the total number of individuals in the entire system, and $\eta_j$ are the usual delta-correlated Gaussian noise sources with zero mean and unit variance.

Applying this scheme to the pairwise interaction model, we set $i=1$, which represents the proportion of ants moving in the direction of food source $1$.  The first, so-called \emph{deterministic}, term of the Langevin equation for $d x_1/dt$, is found to be 
\begin{eqnarray}
\sum_{j=1}^r \nu_{j1} a_j(\boldsymbol{x}) 
& = & \nu_{11} a_1 + \nu_{21} a_2 + \nu_{31} a_3 + \nu_{41} a_4, \nonumber \\
& = & -c x_1 x_2 + c x_1 x_2 - s x_1 + s x_2, \nonumber \\
& = & s (x_2 - x_1). \nonumber
\end{eqnarray} 
Likewise, the second, \emph{stochastic}, term can be computed as
\begin{eqnarray}
\sum_{j=1}^r \nu_{j1}[a_j(\boldsymbol{x})]^{1/2}\mathcal{\eta}_j(t) 
& = & \nu_{11} \sqrt{a_1} \eta_1(t) + \nu_{21} \sqrt{a_2} \eta_2(t) + \nu_{31} \sqrt{a_3} \eta_3(t) + \nu_{41} \sqrt{a_4} \eta_4(t), \\
& = & -\sqrt[]{c\:x_1x_2}\eta_1(t) + \sqrt[]{c\:x_1x_2} \eta_2(t) - \sqrt[]{s\:x_1}\eta_3(t) + \sqrt[]{s\:x_2} \eta_4(t). 
\end{eqnarray} 
Putting these two expressions together, the CLE for the dynamics of the variable $x_1$ is given by
\begin{equation}
\frac{d x_1}{dt} = s\:(x_2 - \:x_1) + \frac{1}{\sqrt[]{N}}\big[\:-\sqrt[]{c\:x_1x_2}\eta_1(t) + \sqrt[]{c\:x_1x_2} \eta_2(t) - \sqrt[]{s\:x_1}\eta_3(t) + \sqrt[]{s\:x_2} \eta_4(t) \:\big].
\label{eq:pairwiseChemicalLangevin}
\end{equation}
It is easy to see that the analogous expression for $x_2$ satisfies $dx_2/dt = - d x_1/dt$, which is consistent with the constraint $x_1 + x_2 = 1$. To obtain the equation for the order parameter $m$, we note that $m = x_1 - x_2$ and find
\begin{equation}
\frac{dm}{dt} = -2sm + \frac{2}{\sqrt[]{N}}\left[ \:\sqrt[]{\frac{c(1-m^2)}{4}}(\eta_2(t) - \eta_1(t)) - \sqrt[]{s \frac{1+m}{2}}\eta_3(t) + \sqrt[]{s \frac{1-m}{2}}\eta_4(t)\:\right].
\end{equation}
By substituting c = 1, rescaling time $t' \to 2st$ and writing $t'=t$, we get
\begin{equation}
\frac{dm}{dt} = -m + \sqrt[]{\frac{N_c}{N}}\left[\sqrt[]{1-m^2}\eta_1'(t) - \sqrt[]{s(1+m)}\eta_2'(t) + \sqrt[]{s(1-m)}\eta_3'(t)\right].
\label{eq:chemical-langevin-1d-pairwise}
\end{equation}
Since the noise sources are Gaussian and independent, the term in the square brackets is equivalent to a single noise source with a variance equal to the $\ell^2-$norm of the variances of the individual noise sources, $\eta_1'$, $\eta_2'$ and $\eta_3'$.  The result is that
\begin{equation}\label{eq:ant-model-langevin-chemical}
\frac{dm}{dt} = -m + \sqrt[]{\frac{N_c}{N}}\,\sqrt[]{1-m^2+2s}\,\eta(t),
\end{equation}
where $N_c = 1/s$. This is the same stochastic differential equation that we derived using the system-size expansion approach~\eqref{eq:ant-model-langevin-syssize}, thus both approximations/methods yield the same mesoscopic description in this case. In the Discussion section, we comment on how the two methods are related, and how they do not necessarily yield the same equations for the extended version of an pairwise interaction model in two spatial dimensions--- {\it i.e.}, with more than two choices--- as set out in~\ref{app:2d-pairwise}. 

\subsection{Characterising mesoscopic dynamics}
\label{sec:solving-coarse-grained-1D}
In this section, we employ three different methods to characterise the previously obtained mesoscopic SDE describing the dynamical behavior of the order parameter, $m$ [Eqs.~\eqref{eq:ant-model-langevin-syssize} or~\eqref{eq:ant-model-langevin-chemical}].
Before doing so, we make some remarks on the qualitative features of the equation. 

First, in the macroscopic ($N\to\infty$) limit, the system exhibits deterministic dynamics given by $dm/dt=-m$. This implies that order decays exponentially with time, reaching the {\it deterministic stable state} $m^*=0$ asymptotically. In this state, an equal number of individuals choose food sources $1$ and $2$. Hence, the macroscopic system is disordered with no consensus or alignment among individuals. 

Second, in the mesoscopic description ($N$ large but finite) both the deterministic and stochastic terms of Eq.~\eqref{eq:ant-model-langevin-syssize} are required to govern the dynamics of $m$. Here, the noise prefactor depends on the current state, $m$; such a feature is called multiplicative or state-dependent noise. In this case, the noise strength is maximum when the system is disordered ({\it i.e.}, $m=0$) and minimum when system is ordered ({\it i.e.}, $m=\pm 1$). Furthermore, the strength of this noise becomes larger for decreasing system sizes.

As we show below, the interplay between a deterministic term that pulls the system towards disorder and a stochastic term that does the opposite can produce interesting and often counter-intuitive dynamics.

\subsubsection{Steady-state solution of the Fokker-Planck equation}
\label{sec:fokker-planck-solution-1D}
SDEs describe how to realise trajectories of stochastic variable(s).  Whilst often preferable for the purposes of gaining heuristic insight, or numerical implementation, they are equivalent to equations of the Fokker-Planck type, which are {\it partial} differential equations (PDEs) for the time-dependent probability distribution that governs the behaviour of such stochastic variable(s). For example, for an SDE of the form $dm/dt = f(m) + g(m) \eta(t)$ (understood according to the It\^{o} convention), the corresponding Fokker-Planck equation is given by~\citep{gardiner2009}
\begin{equation}
\frac{\partial P(m,t)}{\partial t} = - \frac{\partial }{\partial m} \left[f(m) P(m,t)\right] + \frac{1}{2}\frac{\partial^2 }{\partial m^2} \left[g^2(m) P(m,t)\right].
\label{eq:fokker-Planck-f-and-g}
\end{equation}
In most cases, it is not possible to solve this equation analytically.  However, it is often possible to solve this equation for the time-independent steady-state.  We set $\partial P(m,t)/\partial t = 0$ and denote the steady state PDF by $P_s(m)$, giving
\begin{equation}
\frac{\partial }{\partial m} \left[f(m) P_s(m)\right] = \frac{1}{2}\frac{\partial^2 }{\partial m^2} \left[g^2(m) P_s(m)\right].
\end{equation}
Integrating both sides with respect to $m$, we get
\begin{equation}
f(m) P_s(m) = \frac{1}{2}\frac{\partial }{\partial m} \left[g^2(m) P_s(m)\right] + c_0,
\end{equation}
where $c_0$ is the integration constant. Under the assumption of no-flux boundary conditions, $c_0 = 0$. We integrate the above equation again and find that
\begin{equation}
P_s(m) = \frac{1}{P_0} \exp \Big[2 \int_{m0}^m \frac{f(m') - g(m') g'(m')}{g^2(m')} dm' \Big],
\label{eq:Fokker-Planck-solution}
\end{equation}
where $P_0$ is the normalization constant.  Applying this to the mesoscopic description of the pairwise interaction model~\eqref{eq:ant-model-langevin-syssize} where $f(m)=-m$ and $g(m) =\sqrt[]{N_c/N}\sqrt{1+2s-m^2}$, we get
\begin{equation}
\frac{\partial P(m,t)}{\partial t} = \frac{\partial}{\partial m} \left[mP(m,t)\right] + \frac{N_c}{2N} \frac{\partial^2}{\partial m^2}\left[(1+2s-m^2) P(m,t)\right],
\label{eq:Fokker-PlanckOneD}
\end{equation}
whose steady-state solution is given by~\citep{biancalani2014prl}
\begin{equation}
P_s(m) = \frac{1}{\mathcal{P}_0}\frac{1}{(1+2s-m^2)^{1-N/N_c}}, 
\end{equation}
where $N_c = 1/s$ and $\mathcal{P}_0$ is the normalisation constant. In Figure~\ref{fig:numericalSol_SDEVsSSA}, we plot the steady-state distribution as a function of $N$. 

\subsubsection{Numerical integration of the mesoscopic SDE}
\label{sec:NumericalSolSDE_ant}
Realisations of the stochastic trajectories of $m(t)$, governed by $P(m,t)$, can be obtained most easily by numerical integration of the SDE \eqref{eq:ant-model-langevin-syssize}. There exist more complicated schemes, but here we choose an initial value $m(t=0)=m_0$ and employ a simlple Euler-Murayama approach~\citep{press1996numerical}, which corresponds to the difference equation 

\begin{equation}
\Delta m = - m \;\Delta t +  \;\sqrt[]{\frac{N_c}{N}}\;\sqrt[]{1+2s-m^2} \,\sqrt{\Delta t}\,\eta(t),
\end{equation}

\noindent where $\Delta t$ is a small fixed time interval.  In the cases where $P(m,t)$ cannot be obtained from the Fokker-Planck equation, we may repeat the above process a large number of times with different initial conditions and then plot the normalized histogram of $m$.  The result converges, at long times, to the steady-state $P_s(m)$.  In Figure \ref{fig:numericalSol_SDEVsSSA}, we display the time series and the PDF of $m$ for different values of system size $N$.

\subsubsection{Gillespie simulations of the microscopic model}
\label{sec:gillespie-1D}
In both the system-size expansion and the chemical Langevin approach, we assume finite but large systems for deriving Fokker-Planck and hence SDEs. To verify these approaches, we employ stochastic simulations~\citep{gillespie1976,gillespie1977} that are an exact representation of the master equation which describes the underlying microscopic process.  Whilst Figure~\ref{fig:numericalSol_SDEVsSSA} presents results using the original scheme presented by Gillespie, there have since been many studies that improve both the speed and accuracy of the method. Interested readers are referred to studies such as~\cite{erban2009stochastic}, which concerns improving the accuracy of the algorithm, and methods like $\tau$-leaping~\citep{gillespie2001approximate} for accelerating the speed of simulations.

\subsection{Results for the pairwise interaction model}
We now present the results from all three approaches in order to understand both the dynamics and steady-state properties of the order parameter for the pairwise interaction model.
\definecolor{blue}{rgb}{0,0.4470,0.7410}
\definecolor{orange}{rgb}{0.8500, 0.3250, 0.0980}

\newcommand{\blueline}{\raisebox{2pt}{\tikz{\draw[-,blue,solid,line width = 1.5pt](0,0) -- (5mm,0);}}}

\newcommand{\orangeline}{\raisebox{2pt}{\tikz{\draw[-,orange,solid,line width = 1.5pt](0,0) -- (5mm,0);}}}

\newcommand{\blackline}{\raisebox{2pt}{\tikz{\draw[-,black,solid,line width = 1.5pt](0,0) -- (5mm,0);}}}

\begin{figure}[!]
\begin{adjustbox}{addcode={\begin{minipage}{\width}}{\caption{Dynamics and steady-state properties of the order-parameter or collective behaviour ($m$) for the pairwise interaction model. Top row: Blue line (\protect\blueline) indicates numerical integration of the analytically derived SDE equation~\eqref{eq:ant-model-langevin-syssize} whereas the orange line (\protect\orangeline) shows Gillespie simulations of the model. Bottom row: Steady-state distribution of the order-parameter $m$ obtained by numerical integration of SDE, Gillespie simulations and the analytical solutions of the Fokker-Planck equation for the model are shown by the black line (\protect\blackline). $N_c = 250$
\label{fig:numericalSol_SDEVsSSA}
      }\end{minipage}},rotate=90,center}
\includegraphics[scale = 0.78]{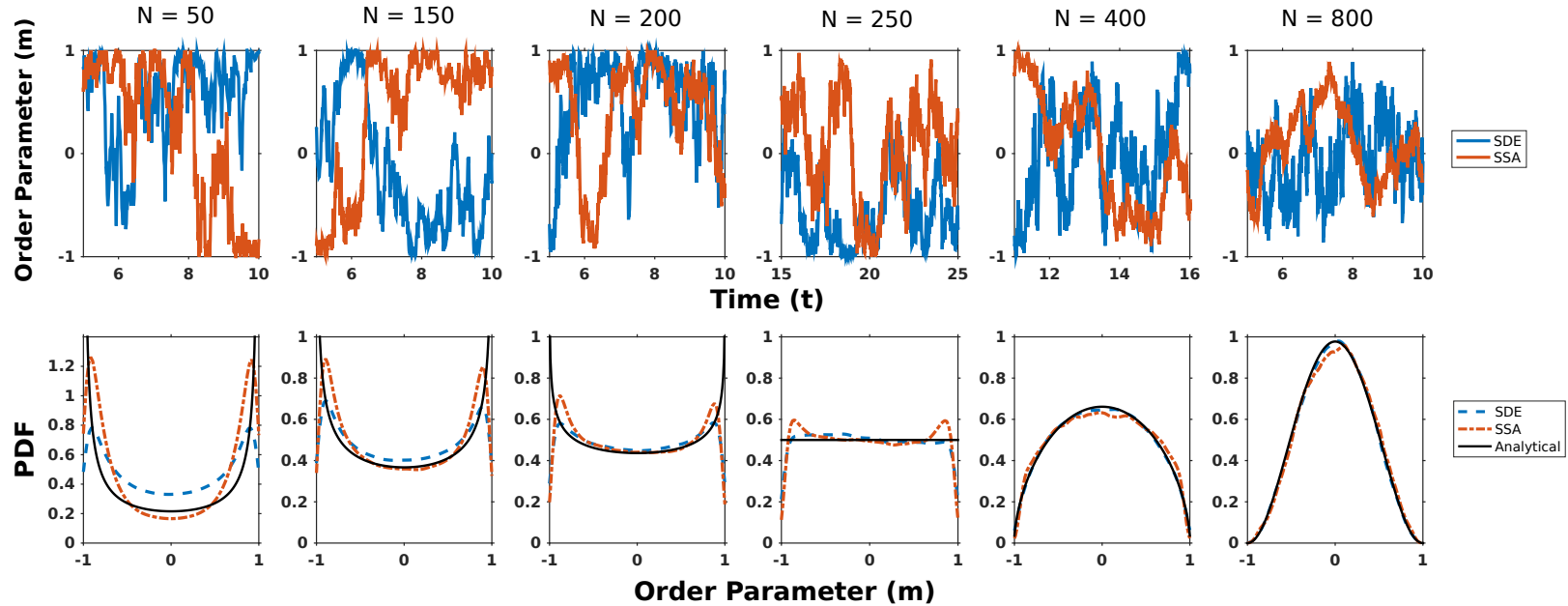}
\end{adjustbox}
\end{figure}

The steady state solution of the Fokker-Planck equation predicts a group-size dependent PDF of the order parameter $m$ (bottom row of Figure~\ref{fig:numericalSol_SDEVsSSA}). Specifically, for large group sizes ($N>N_c$) the PDF $P_s(m)$ is unimodal with the mode occurring at $m=0$. This implies that the most likely state is a disordered state consistent with the deterministic fixed point of the macroscopic/deterministic description. By contrast, for small group sizes $N<N_c$, the steady-state PDF $P_s(m)$ is {\it bi}modal, with modes at $m=\pm 1$. Therefore, the system spends most of its time in highly ordered states. These two states correspond to a consensus decision that results in clear collective movement of individuals in a single direction.

The PDF obtained from repeated numerical integration of the SDE~\eqref{eq:ant-model-langevin-syssize} produces qualitatively similar results as the analytical solution of the Fokker-Planck equation indicating minimal transient dynamics before reaching the steady state. The advantage of numerical simulation of the SDE is that we can also visualise the time series of the order parameter $m$ (\blueline $\,$ in top panel, in Figure \ref{fig:numericalSol_SDEVsSSA}). We find that, for smaller group sizes, $m$ frequently switches back and forth between two ordered states at $m=1$ and $m=-1$.  However, this is clearly not the case with larger groups, which spend large amounts of time in a disordered $m=0$ state.

As expected, both the steady-state solution to the Fokker-Planck equation and repeated numerical integration of the SDE qualitatively match with the Gillespie simulations. Of course, as system size decreases, the results diverge slightly, which highlights the fact that the mesoscopic descriptions are only approximations of the underlying microscopic behaviour.

\section{Ternary interaction model for binary choice}\label{sec:higher-order-locustmodel-1d}

We now present a simple extension of the pairwise interaction model that includes higher-order interactions. This model of collective behaviour was originally developed by~\cite{dyson2015} to explain the quantitative behaviour of a march of locusts in a one-dimensional arena~\citep{buhl2006} and incorporates interactions between three individuals in addition to the previously discussed pairwise interactions; we refer to this model as the `ternary interaction model'. Once again, we apply both the system-size expansion and chemical Langevin approaches to derive a mesoscopic description. 

As in the pairwise model, individuals move towards food source $1$ or $2$ depending on both their interactions with others and random factors. We once again assume that individuals spontaneously switch at a rate $s$ between food sources/directions  
\begin{subequations}\label{eq:stochastic-HigherOrder}
\begin{equation}
X_1 \xrightarrow {s}  X_2, 
\label{eq:stochasticA-HigherOrder}
\end{equation}
\begin{equation}
X_2 \xrightarrow {s}  X_1.
\label{eq:stochasticB-HigherOrder}
\end{equation}
\end{subequations}
Likewise, we also retain the copying interactions, occurring at a rate $c$, of the pairwise interaction model, given by
\begin{subequations}\label{eq:copying-HigherOrder}
\begin{equation}
X_1 + X_2 \xrightarrow {c} 2 X_1, 
\label{eq:copyingA-HigherOrder}
\end{equation}
\begin{equation}
X_2 + X_1 \xrightarrow {c} 2 X_2.
\label{eq:copyingB-HigherOrder}
\end{equation}
\end{subequations}
The new component of this model is a higher-order interaction, where three individuals interact.   The reaction occurs at a specific rate $h$, and results in all individuals moving in the direction of the majority. Depending on whether $X_1$ or $X_2$ was in the majority at the time of interactions, one of the two following reactions occur:
\begin{subequations}\label{eq:higherOrder}
\begin{equation}
2X_1 + X_2 \xrightarrow {h} 3 X_1, 
\label{eq:higherOrderA}
\end{equation}
\begin{equation}
2X_2 + X_1 \xrightarrow {h} 3 X_2.
\label{eq:higherOrderB}
\end{equation}
\end{subequations}
\noindent We note that \cite{schulze2005monte} use a similar model to study evolution of language. In that model, a focal individual modifies its language spontaneously at a rate $p$, copies the language of a randomly chosen individual at a rate $q$, and finally, with a probability $(1-x)^2r$, the individual chooses the the language spoken by the fraction $x$ of the population. This last reaction is identical to the ternary interaction set out in Eqs.~\eqref{eq:higherOrder}. 

\subsection{Constructing a mesoscopic SDE for the ternary interaction model}
\label{sec:coarse-grained-equations-higher-order}
As we did before with the pairwise interaction model, we now apply both the system-size expansion and the chemical Langevin equation methods in order to derive a mesoscopic equation for the order parameter $m=x_1-x_2$.

\subsubsection{van Kampen's system-size expansion of transition rates}
\label{sec:coarse-grained-derivation-locust}

\textit{Transition Rates}: The state variables of the system are again $x_1$ and $x_2$: the proportion of individuals moving in the direction of food sources 1 and 2, respectively. Similarly, since $x_1+x_2 = 1$, one of either $x_1$ or $x_2$ can completely describe the state of the system. The transition rates $T_{21}$ and $T_{12}$ are the same as in the pairwise interaction model, except they must also incorporate reactions \eqref{eq:higherOrderA} and \eqref{eq:higherOrderB}.  The additional rates follow the `mass action' rationale set out previously, giving
\begin{equation}
T_{21} = T\left(x_1-\frac{1}{N},x_2+\frac{1}{N} \bigg\vert \,x_1,x_2\right) = sx_1 + cx_1x_2 + hx_1x_2^2,
\end{equation}
and 
\begin{equation}
T_{12} = T\left(x_1+\frac{1}{N},x_2-\frac{1}{N} \bigg\vert \,x_1,x_2\right) = sx_2 + cx_1x_2 + hx_1^2x_2.
\end{equation}

\textit{Fokker-Planck equation}:
Next, we substitute these transition rates into the general form of the Fokker-Planck equation for binary-choice models \eqref{eq:fokker-Planck-transition-rates12}, where we note that the steps from Eq.~\eqref{eq:masterEquation} to Eq.~\eqref{eq:fokker-Planck-transition-rates12} are independent of model interaction details. Therefore, we get
\begin{equation}
\begin{split}
\frac{\partial P(x_1,x_2,t)}{\partial t} = & \left[ - \sum_{i=1}^2\frac{\partial [\mathcal{A}_i\:P(x_1,x_2,t)]}{\partial x_i} + \frac{1}{2N}\sum_{i,j=1}^2\frac{\partial^2\: [\mathcal{B}_{ij}\:P(x_1,x_2,t)]}{\partial x_ix_j}\right], \\
\end{split}
\label{eq:fokkerPlanck-higher-order}
\end{equation}
where,
\begin{eqnarray}
\mathcal{A}_1 & = & T_{12} - T_{21} = s(x_2-x_1) + hx_1x_2(x_1-x_2), \\
\mathcal{A}_2 & = & - \mathcal{A}_1, \\
\mathcal{B}_{ij} & = & (T_{ij}+T_{ji})(-1)^{i+j} = (-1)^{i+j}\left[s(x_1+x_2)+2cx_1x_2+hx_1x_2(x_1+x_2)\right].
\end{eqnarray}

\textit{Mesoscopic SDE}: The coupled SDEs that correspond to realisations of the above Fokker-Planck equation are given by
\begin{equation}
\frac{dx_1}{dt} = \mathcal{A}_1+\frac{1}{\sqrt[]{N}}\xi_1(t),\ \mathrm{and}\ \frac{dx_2}{dt} = \mathcal{A}_2+\frac{1}{\sqrt[]{N}}\xi_2(t),\ \mathrm{where}\ \langle\xi_i(t)\xi_j(t')\rangle = \mathcal{B}_{ij} \delta(t-t').
\label{eq:langevin-higher-order-sys-size}
\end{equation}
Using the same set of manipulations that we followed while analysing the pairwise interaction model, we obtain the mesoscopic equation for the order parameter $m$ for the ternary interaction model. To do so, we use the transformation: $\xi_i = \sum_{j=1}^2\mathcal{G}_{ij}\eta_j$ , such that $\mathcal{G}$ satisfies $\mathcal{B} = \mathcal{G}\mathcal{G}^T$. We find that $$\mathcal{G}_{ij}\\ = (-1)^{i+j+1}\sqrt[]{s(x_1+x_2)+2cx_1x_2+hx_1x_2(x_1+x_2)}/\sqrt[]{2}.$$After further transforming to $m=x_1-x_2$, we obtain the mesoscopic dynamical equation~\citep{dyson2015}

\begin{equation}
\frac{dm}{dt} = -2sm+\frac{h}{2}m(1-m^2) + \frac{2}{\sqrt[]{N}}\sqrt[]{s+\frac{2c+h}{4}(1-m^2)}\,\eta(t),
\label{eq:langevin-higher-order}
\end{equation}
which, under the time rescaling $t'=2st$, recovers the mesoscopic equation of the pairwise interaction model on setting $h=0$.

\subsubsection{Chemical Langevin approach}
As in Section \ref{sec:chem-lagenvin-ant-model-1d}, we define both propensity functions $a_j(\boldsymbol{x})$ and the state change, or stoichiometry, vector $\boldsymbol{\nu}$ for the ternary interaction model (Table~\ref{tab:chemicalLangevin-HigherOrder}). To derive the relevant ternary mesoscopic equation, we substitute both $a_j(\boldsymbol{x})$ and $\boldsymbol{\nu}$ from Table~\ref{tab:chemicalLangevin-HigherOrder} into the chemical Langevin equations,    
\begin{equation}
\frac{d x_i(t)}{dt} = \sum_{j=1}^r \nu_{ji} a_j(\boldsymbol{x})  + \frac{1}{\sqrt[]{N}}\sum_{j=1}^r \nu_{ji}[a_j(\boldsymbol{x})]^{1/2}\mathcal{\eta}_j(t),\ \textrm{for}\; i = 1,...,n,
\label{eq:chemical-langevin-general-higher-order}
\end{equation}
where $n$ is the number of species, $r$ is the number of reactions, $N$ is the total number of individuals, and $\eta_j$ are Gaussian white noise sources with zero mean and correlation $\langle\eta_i(t)\eta_j(t^\prime)\rangle = \delta_{ij}\delta(t-t^\prime)$.

\begin{table}[h]
\centering
\begin{tabular}{|c|c|c|r|r|} 
\hline   
S. No. (j)& Reaction & Propensity & $\boldsymbol{\nu}_{j1}$ & $\boldsymbol{\nu}_{j2}$ \\ [0.5ex]
\hline  
1&$X_1 \xrightarrow {s}  X_2$   &$sx_1$ &-1 &1\\
2&$X_2 \xrightarrow {s}  X_1$   &$sx_2$ &1 &-1\\
3&$X_1 + X_2\xrightarrow {c}  2X_1$   &$cx_1x_2$ &1 &-1\\
4&$X_1 + X_2\xrightarrow {c}  2X_2$   &$cx_1x_2$ &-1 &1\\
5&$2X_1 + X_2\xrightarrow {h}  3X_1$   &$hx_1^2x_2$ &1 &-1\\
6&$2X_2 + X_1\xrightarrow {h}  3X_2$   &$hx_1x_2^2$ &-1 &1\\
[1ex] 
\hline                          
\end{tabular}
\caption{Reactions in the higher order interaction model, their propensities and the state change vector}
\label{tab:chemicalLangevin-HigherOrder}
\end{table}
\noindent As an example, consider the variable $x_1$.  The deterministic term of the CLE is given by
\begin{equation}
\begin{split}
\sum_{j=1}^r \nu_{j1} a_j(\boldsymbol{x})
& = \nu_{11} a_1 + \nu_{21} a_2 + \nu_{31} a_3 + \nu_{41} a_4 + \nu_{51} a_5 + \nu_{61} a_6 \\
& = -sx_1 + sx_2 + cx_1x_2 - cx_1x_2 + hx_1^2x_2 - hx_1x_2^2 \nonumber\\
& = s(x_2 - x_1) + hx_1x_2(x_1-x_2).
\end{split}
\end{equation}
The stochastic term is given by
\begin{equation}
\begin{split}
\sum_{j=1}^r \nu_{j1} [a_j(\boldsymbol{x})]^{1/2} 
& = \nu_{11} \sqrt{a_1} \eta_1(t) + \nu_{21} \sqrt{a_2} \eta_2(t) + \nu_{31} \sqrt{a_3} \eta_3(t) + \nu_{41} \sqrt{a_4} \eta_4(t) \\
& \quad + \nu_{51} \sqrt{a_5} \eta_5(t) + \nu_{61} \sqrt{a_6} \eta_6(t)\nonumber\\
&= -\sqrt[]{sx_1}\eta_1 + \sqrt[]{sx_2}\eta_2 + \sqrt[]{cx_1x_2}\eta_3 - \sqrt[]{cx_1x_2}\eta_4 + \sqrt[]{hx_1^2x_2}\eta_5 - \sqrt[]{hx_1x_2^2}\eta_6.
\end{split}
\end{equation}
As before, it is easy to show that $dx_1/dt = -dx_2/dt$, because $\nu_{j1} = -\nu_{j2}$, for all $j$. As noise sources $\eta_1$ to $\eta_6$ are independent we may combine them into a single noise source to obtain
\begin{equation}
\sum_{j=1}^r \nu_{j1} [a_j(\boldsymbol{x})]^{1/2} = \sqrt[]{s(x_1+x_2)+2cx_1x_2+hx_1x_2(x_1+x_2)}\eta(t). \nonumber
\end{equation}
Hence, the CLE for the higher order interaction model in terms of variable $x_1$ and $x_2$ is given by
\begin{equation}
\frac{dx_1}{dt} = s(x_2 - x_1) + hx_1x_2(x_1-x_2) + \frac{1}{\sqrt[]{N}}\sqrt[]{s(x_1+x_2)+2cx_1x_2+hx_1x_2(x_1+x_2)}\eta(t) = -\frac{dx_2}{dt}.
\end{equation}
Substituting $m = x_1-x_2$ and using the constraint $x_1+x_2=1$, we get
\begin{equation}
\frac{dm}{dt} = -2sm + \frac{h}{2}m\,(1-m^2) + \frac{2}{\sqrt[]{N}}\sqrt[]{s+\frac{2c+h}{4}(1-m^2)}\eta(t).
\end{equation}
This mesoscopic dynamical equation in $m$, for the one dimensional ternary interaction model, derived using the chemical Langevin approach is exactly the same equation derived using the system-size expansion method. Therefore, we again see that both methods give exactly same results for one dimensional/uni-variate models. 

\subsection{Characterising mesoscopic dynamics}
We now turn to characterising the mesoscopic SDE~\eqref{eq:langevin-higher-order} of the ternary interaction model both numerically and analytically. The methods are similar to the pairwise interaction model (Section \ref{sec:solving-coarse-grained-1D}).

\subsubsection{Steady-state solution of the Fokker-Planck equation}
The deterministic and the stochastic terms of the mesoscopic equation for the ternary interaction model are $f(m) = -2s\,m + h\,m(1-m^2)/2,$ and $g(m) = 2\,\sqrt[]{\left(s+(2c+h)(1-m^2)/4\right)/N}$, respectively. Therefore, the Fokker-Planck equation for the ternary interaction model is given by
\begin{equation}
\frac{\partial P(m,t)}{\partial t} = \frac{\partial}{\partial m}\left\{\left[2sm - \frac{h}{2}m(1-m^2)\right]P(m,t)\right\} + \frac{2}{N}\frac{\partial^2}{\partial m^2}\left\{\left[s+\frac{2c+h}{4}(1-m^2)\right]P(m,t)\right\}.
\label{eq:fokker-planck-higher-order}
\end{equation}
Using the general expression for steady state PDF given by Eq.~(\ref{eq:Fokker-Planck-solution}) \citep{gardiner2009}, we get 
\begin{equation}
P_s(m) = \frac{1}{P_0}[4s+(2c+h)(1-m^2)]^{\frac{4Ns(c+h)}{(2c+h)^2}-1}e^{\frac{hm^2N}{2(2c+h)}},
\label{eq:steady-state-solution-higer-order}
\end{equation}
where $P_0$ is the normalization constant \citep{dyson2015}. In Figure \ref{fig:numericalSol_SDEVsSSA_higherOrder}, we plot this expression for different values of $N$.

\subsubsection{Numerical integration of the mesoscopic SDE}
We applied the Euler-Murayama scheme to the SDE~\eqref{eq:langevin-higher-order} in exactly the same way as Section \ref{sec:NumericalSolSDE_ant}. The time series and the probability density function thus obtained are shown in Figure \ref{fig:numericalSol_SDEVsSSA_higherOrder} for different values of $N$.

\subsubsection{Gillespie simulations of the microscopic model}

We implement the Gillespie algorithm in the same way as described in Section~\ref{sec:gillespie-1D}.  We choose the sampling time as $1/s$, independent of system size. We later rescaled the time $t = t'/N$ to match the timescale of the mesoscopic equation~\eqref{eq:langevin-higher-order}. The results of these simulations are presented in Figure \ref{fig:numericalSol_SDEVsSSA_higherOrder}. 

\subsection{Results for the ternary interaction model}
\label{sec:results-higher-prder}

\begin{figure}[!]
\begin{adjustbox}{addcode={\begin{minipage}{\width}}{\caption{\textbf{Dynamics and steady state distributions of order parameter for the higher order interaction model} Top row: Time series of order parameter ($m$) for three group size ($N=50,400,800$) from the numerical integration of the stochastic differential equation (sde) \eqref{eq:langevin-higher-order}, shown in (\protect\blueline) and from the stochastic simulation algorithm (SSA) method (\protect\orangeline). Bottom row: Corresponding steady state probability density functions of $m$ obtained from numerical solutions of sde, SSA and the analytical solution of the Fokker-Planck equation \eqref{eq:fokker-planck-higher-order}, shown in (\protect\blackline). Parameter values: $s = 0.05$, $c = 0.005$, $h = 0.21$  
\label{fig:numericalSol_SDEVsSSA_higherOrder}
      }\end{minipage}},rotate=0,center}
\includegraphics[scale=0.85]{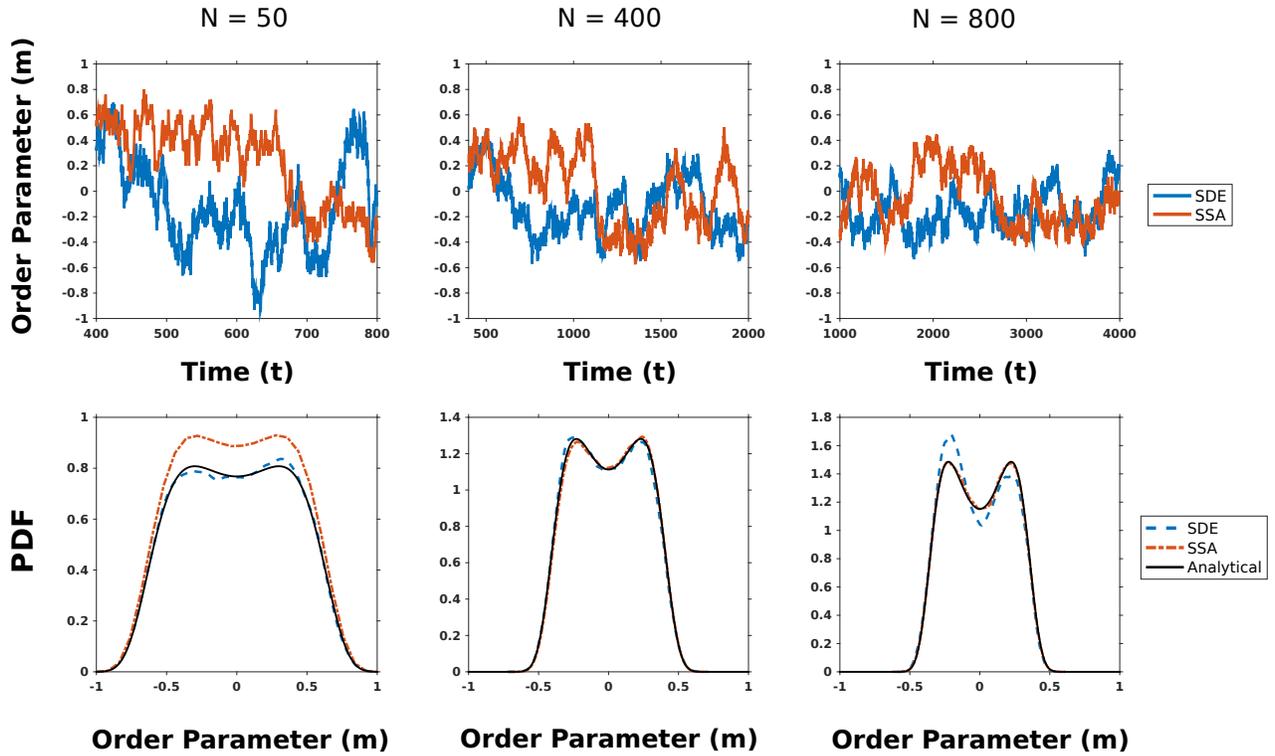}
\end{adjustbox}
\end{figure}

The steady-state solution of the Fokker-Planck equation is bimodal for all group sizes (\blackline $\,$in bottom row, Fig.~\ref{fig:numericalSol_SDEVsSSA_higherOrder}). The two modes of the distribution are however less distinct for small system sizes, becoming more prominent with increasing system size. Increasing the rate of higher-order interactions ($h$) moves the location of the distribution modes to larger values of $m$. This implies that the system size only has quantitative but not qualitative influence on the steady state distribution of the order parameter. The effect is therefore in sharp contrast to the pairwise interaction model, where increasing system size resulted in a transition from bimodal to unimodal distribution of $m$. Specifically, in the pairwise interaction model we observed two clear modes for small systems whereas, for the ternary interaction model, this corresponds to large system sizes.  

To understand these results, consider the deterministic limit of the ternary interaction model's SDE. The $N\to\infty$ limit of equation \eqref{eq:langevin-higher-order} has three fixed points ($0,\pm\sqrt[]{(h-4s)/h}$). For $h>4s$, three real roots exist, of which $m^*=\pm\sqrt[]{(h-4s)/h}$ are stable since $f'(m^*) < 0$. Visual inspection of two modes in the steady state PDF reveals that they are indeed near the location of these two deterministic stable fixed points of the dynamical equation for all values of the system size $N$. This is unlike the pairwise interaction model where the equation has only one stable fixed point at $m=0$, but nevertheless has a bimodal PDF whose modes are away from $m=0$; an effect referred-to as noise-induced bistability~\citep{horsthemke1984}. 

The effect of system size $N$ in the mesoscopic description is captured in the stochastic term of \eqref{eq:langevin-higher-order}, which is $\mathcal{O}\left(1/\sqrt[]{N}\right)$. At small system sizes, the strength of stochasticity is relatively large and therefore, the order parameter is constantly pushed away from the stable equilibria $m^*=\pm\sqrt[]{(h-4s)/h}$. As a result, when $N=50$ we observe a relatively large spread around the two stable states in the distribution of $m$ (see bottom row in Figure \ref{fig:numericalSol_SDEVsSSA_higherOrder}). On the other hand, for large system size the strength of stochasticity is less and thus system resides longer in the stable states. Consequently, the distribution shows two clear modes corresponding to deterministic stable equilibria in the system in low noise/large system size case. 

We can explore the temporal evolution of the  order parameter, $m$, by numerically integrating Eq~\eqref{eq:langevin-higher-order} (\blueline $\,$in Figure \ref{fig:numericalSol_SDEVsSSA_higherOrder}) or by Gillespie simulations (\orangeline $\,$in Figure \ref{fig:numericalSol_SDEVsSSA_higherOrder}). For small group sizes $m$ does not reside evidently in the stable states, moving constantly across different states (top row of Figure \ref{fig:numericalSol_SDEVsSSA_higherOrder}). As the group size increases, the  system resides longer in the two stable states ($m^*=\pm\sqrt[]{\frac{h-4s}{h}}$), exhibiting occasional transitions between them. This effect was further quantified by \cite{dyson2015} by calculating the mean residence time in the two stable states, which increases with group size. The results from all three approaches (the analytical approach, the SDE simulations and the Gillespie simulations) to solving the mesoscopic dynamics of the ternary interaction model qualitatively match with each other, with slight discrepancy for small groups sizes (Bottom row in Figure \ref{fig:numericalSol_SDEVsSSA_higherOrder}). This is not surprising since the Langevin and the Fokker-Planck equations approximates the dynamics for large but finite systems, but Gillespie simulations are exact representation of the underlying process.

\section{Discussion}
In this book chapter we presented two approaches from the literature, the system-size expansion method~\citep{van1992stochastic} and the chemical Langevin equation method \citep{gillespie2000}, to construct mesoscopic dynamical equations of collective behavior from microscopic rules. These methods allow us to capture how, for finite population sizes, stochasticity at the microscopic scale manifests at the mesoscopic scale. The resulting behaviour is captured by stochastic differential equations, sometimes referred to as Langevin equations. We demonstrated applications of these approaches to two simple models of collective decision-making involving individuals that must make a binary choice; in both cases, both methods yield identical Langevin equations. In the following Section~\ref{sec:compare-methods} we therefore compare various technical aspects of the two complementary methods. In Section~\ref{sec:compare-models}, we discuss how the aforementioned techniques reveal important differences regarding the two models we considered; one considers only pairwise interactions between individuals, whilst the other incorporates higher-order interactions.  Finally, in Section~\ref{sec:extensions}, we discuss some extensions of the model. 

\subsection{Comparison of system-size expansion with chemical Langevin approach} \label{sec:compare-methods}

Both methods of constructing coarse-grained descriptions at mesoscopic scales are based on similar sets of assumptions. The system-size expansion method involves a formal procedure that begins with the master equation and then expands its terms in a small quantity: the inverse of the system size, $1/N$. By assuming $N$ large and thereby keeping only two leading order terms, we obtain a Fokker-Planck equation, which describes the temporal evolution of the PDF of the order-parameter, $m$. This is equivalent to assuming that the rate-of-change in the probability distribution for $m$ is fully determined by its first and second jump-moments, and hence that the noise is Gaussian in character. In contrast to this formal approach, the chemical Langevin method is based on heuristics that keep track of both the probability of each `chemical reaction' (or interaction), and how each reaction changes the state of the system. To do this, the time-scale for reactions is so chosen that the changes in the number of individuals of each species can be approximated by a Gaussian/normal distribution. Thus, the two methods effectively make the same set of assumptions about the noise process that governs mesoscopic dynamics of the coarse-variable. [A more detailed discussion is provided in \cite{gillespie2002chemical}]. Given this equivalence, it is not surprising that two methods do indeed yield identical Fokker-Planck equations and hence SDEs for the two univariate models under consideration.

More generally, the system-size expansion yields a Fokker-Planck equation, whilst the chemical Langevin approach results in one or more coupled SDEs.  There is, of course, a formal equivalence between Fokker-Planck equations and SDEs. However, the decomposition of a Fokker-Planck equation into SDEs is not unique for multivariate systems.  Specifically, this requires the decomposition of a known matrix--- the matrix of second jump moments--- into the product of an unknown matrix and its transpose. There exist many methods for such a decomposition, the Cholesky decomposition being a popular choice that ensures the unknown matrix is square but requires that the matrix of second jump moments is positive definite.  In this case, the result is a set of coupled SDEs that involve the minimum number of possible noise sources (equal to the number of independent degrees-of-freedom required to fully describe the state of the system).  By contrast, the chemical Langevin approach results in a system of coupled SDEs which involve as many noise sources as there are `chemical reactions'. Typically, the fewer noise terms of the Cholesky decomposition involve cumbersome and hard-to-simplify prefactors, whilst the many terms of the chemical Langevin equations have relatively straightforward prefactors.  We demonstrate this point in~\ref{app:2d-pairwise} where we consider an extension of the binary-choice to four-choices, representing collective movement in two-dimensions rather than one. Using the system-size expansion to obtain a Fokker-Planck equation and then employing a Cholesky decomposition results in SDEs that are, for all intents and purposes, intractable, due to the complexity of the noise prefactors.  However, by comparison, the chemical Langevin equations are significantly easier to construct and analyze.

We nevertheless stress that the two descriptions, as presented, \emph{are} statistically the same, since they correspond to the same Fokker-Plack equation.  The difference is purely aesthetic, and arises in multivariate systems due to the non-uniqueness of mapping a Fokker-Planck equation to a set of SDEs.  However, one area, so far not discussed, where the two approaches do differ is the so-called Linear Noise Approximation (LNA)~\citep{van1992stochastic}. The LNA is attributed to van Kampen and typically used in conjunction with the expansion of transition rates in inverse system-size.  It recognises that, formally, expansions of this type should be accompanied by an ansatz regarding the $N$-dependence of the underlying variables.  If the variance of the underlying variables is expected be proportional to $\sqrt{N}$, then this affects how the various terms of the expansion equate at lowest order, and results in mesoscopic descriptions that are \emph{additive} in noise, rather than multiplicative.  By contrast, if the variance is not expected to be Gaussian-like, then it does not affect the expansion at lowest order, and hence the LNA is not required.

\subsection{Multiplicative noise at mesoscopic scales} \label{sec:compare-models}
To better understand the mesoscopic descriptions that result from the aforementioned approaches, we considered two simple models of collective behaviour from the literature, where individuals must make a binary choice between two alternatives. The models differ only in the way individuals interact. In the first case we assumed pairwise interactions \citep{kirman1993quartecon,biancalani2014prl,lux1995herd}, whilst in the second case three individuals were permitted to interact at any given time~\citep{dyson2015}. In the literature, \cite{biancalani2014prl} and \cite{dyson2015} used the system-size expansion approach to derive mesoscopic equations for such models. Here, we recapitulate their results and, in addition, demonstrate the chemical Langevin equation approach; the latter having been used extensively for analyzing chemical reactions and biochemical networks, but much less in the ecology literature~\citep{datta2010}.

From this exercise, we may learn a number of interesting points about noise at mesoscopic scales.  First and foremost, for the simple binary-choice models considered, the SDEs for the coarse-variable (consensus/order) contain multiplicative noise, {\it i.e.}, where the strength of stochasticity depends on the current state of the system. In other words, irrespective of the method of derivation, the resulting SDE is of the form $dm/dt = f(m) + g(m) \eta(t)$, with $g(m) \ne$ constant. Secondly, studying these two models (both pairwise and ternary interactions) together provided an opportunity to highlight how the nature of interactions can subtly influence collective behaviour at mesoscopic scales. In both models, the noise term $g(m)$ decreases with group size by $\mathcal{O}(1/\sqrt{N})$, however the effect of noise on collective behaviour contrasts sharply between the two. Perhaps counter-intuitively, increasing the strength of noise in the SDE of the pairwise interaction model, actually increases order, despite the ostensibly randomising effects of noise.  This is exactly the opposite for the ternary interaction model, where increasing noise reduces the order, or level of consensus, among individuals. 

To understand this quantitatively, consider the deterministic limit  ($N\to\infty$), in which $dm/dt = f(m)$.  Here, we expect the system to asymptotically reach one of the deterministic stable states, $m^*$, given by $f(m^*)=0$ and $f'(m^*)<0$. However, when the system is driven by multiplicative noise we have $dm/dt = f(m) + g(m) \eta(t)$, where the form of $g(m)$ determines the most probable states of the system~\citep{horsthemke1984}.  Crucially, these do \emph{not} necessarily correspond to the deterministic stable states, as they would if the noise were simply additive [i.e., $g(m)=\mathrm{constant}$].  This is precisely the case for the pairwise interaction model; the deterministic stable state corresponds to $m^*=0$, but the most likely states ({\it i.e.}, the modes of the distribution) are close to $m^* \pm 1$. Thus, the multiplicative noise not only moves the system away from its deterministic stable state but also increases order. By contrast, in the higher-order (ternary) interaction model, the most likely states are indeed around or close to deterministic stable states. 

\subsection{Extensions and concluding remarks}\label{sec:extensions}

In this chapter, we focussed only on two simple models of collective behaviour where interactions rules were either pairwise or at most ternary. Further, we focussed on non-spatial models and assumed that the system is fully-connected {\it i.e.} any individual can interact with any other individual. We deliberately chose these simple models because the main purpose of this chapter was to illustrate two complementary mathematical methods to deriving mesoscopic dynamics of the collective behaviour; elucidating how stochsaticity arising from individual behaviours is amplified at mesoscopic scales by small group sizes. 

An obvious extension of this work includes considering more complicated higher-order interactions; many empirical studies quantify nature of interactions suggest that interactions\citep{biro2006compromise,katz2011inferring,herbert2011inferring,bialek2012statistical,gautrais2012deciphering,mann2013multi,jiang2017Idetifying} with some suggesting that interactions among flock-members can extend to several members, for example up to seven in starling flocks~\citep{ballerini2008interaction}. For better representation of realistic flock dynamics, a continuous state/direction, which is in contrast to discrete states we have considered so far, could be incorporated~\citep{othmer1988models,gavagnin2018stochastic}. However, analysis of continuous state models may require more sophisticated analytical tools than presented in this chapter. Further, the emergence of collective behavior at mesoscopic scales, which is absent at macroscopic scales, is also of relevance to understand the evolution of collective behaviour. Much of the previous work concerning the evolution of social behaviour in animals has involved either deterministic game theory, which assumes infinite population sizes~\citep{krause2002living,sumpter2010principles,torney2010specialization}, or simulations of large numbers of particles~\citep{reluga2005simulated,spector2005emergence,wood2007evolving,guttal2010,guttal2012cannibalism}. Recent studies have indeed begun to highlight the role of stochasticity arising from small group sizes and finite population sizes~\citep{joshi2017mobility,joshi2018Demographic} where mesoscopic descriptions are important. 

Further, our assumption of fully-connected systems may be valid for small groups of animals, which is the focus of our paper, but for large groups, accounting for the local nature of interactions becomes important. Mesoscopic descriptions of such large but finite systems have been described for a small number of cases, and result in stochastic partial differential equations~\citep{dean1996langevin,biancalani2010stochastic,mckane2014stochastic}. On the other hand, the macroscopic or hydrodynamic descriptions of the collective motion models discussed in the first section have been well studied~\citep{toner1995long,ramaswamy2010annrev,marchetti2013rmp}. It would therefore be interesting to understand how mesoscopic descriptions, which focus on capturing stochasticity arising from finite sized systems, dovetail with hydrodynamic descriptions at ostensibly larger length scales.

In summary, we presented two complementary approaches to derive mesoscopic descriptions of simple collective behaviour models. It is worth noting, however, that the methods we presented are applicable to any ecological model that are based on asynchronous stochastic update rules and where spatial structure is not important. For example, these methods can also be useful to deriving mesoscopic description of population and community dynamics which captures demographic stochasticity~\citep{sabiha2018thesis}. Whilst van Kampen's system-size expansion has indeed been applied to study simple ecological and evolutionary scenarios~\citep{mckane2005prl,traulsen2012,black2012stochastic}, we have highlighted the difficulty of employing this method for studying multi-species interactions. In such cases, we suggest that the method of chemical Langevin approach could be powerful. Therefore, we hope that our book chapter provides a pedagogical review of mathematical methods for describing mesoscopic dynamics that are useful not only for collective behaviour models but also for other biological dynamics where finite sizes of populations/groups is important.

\section{Resources}
The codes used for simulating the pairwise interaction and the ternary interaction models and the corresponding mesoscopic stochastic differential equations can be found at GitHub (\url{https://tinyurl.com/y8zvauqx}).

\section{Acknowledgements}
We thank Sabiha Majumder, Sumithra Sankaran, Jaideep Joshi, Jeffrey Phillippson, and Appilineni Kushal for various insightful discussions. We are thankful to Shubham Rana for useful comments on the manuscript. JJ acknowledges support by the Council for Scientific and Industrial Research, India for research scholarship. RGM acknowledges support from Simons Foundation. VG acknowledges support from DBT-IISc partnership program, Science and Engineering Research Board (DST, Govt of India) and DST-FIST for infrastructure support. 

\appendix
\section{The Chemical Langevin equation}\label{app:chemlangevin} 

In this appendix, we briefly recapitulate the approach developed by~\cite{gillespie2000}. We consider $n$ chemicals/species $\{X_1,...,X_n\}$ whose interactions are represented via $r$ chemical reactions $\{R_1,...,R_r\}$. We define the state of the system by $\boldsymbol{N}(t)$ $\equiv$ $\{N_1(t),...,N_n(t)\}^\mathsf{T}$ where $N_i(t)$ denotes the number of $X_i$ molecules in the system at time $t$. Next, we define a propensity function $a_j(\boldsymbol{N}(t))$, which represents the probability that for a given state $\boldsymbol{x}(t) = \boldsymbol{N}(t)/N$, the reaction $R_j$ will occur within the next infinitesimal time interval $t + dt$. Here, $N = \sum_i N_i$ is the total number of individuals in the system. We also define the state change matrix $\boldsymbol{\nu}_j$ whose \textit{i}th component $\boldmath{\nu} _{ji}$ is defined by the change in the number of $X_i$ molecules due to the occurrence of reaction $R_j$. For example, consider the reaction $R_1:X_1 + X_2 \rightarrow 2X_1$ and the reverse reaction $R_2: 2X_1 \rightarrow X_1 + X_2$. The propensity function for the reaction $R_1$ would be $c_1 N_1 N_2/N^2$, and for the reaction $R_2$ it would be $c_2N_1(N_1-1)/2N^2$. The state change matrix would be $\boldsymbol{\nu}_1 = (+1,-1,0,...,0)$, such that $\boldsymbol{\nu}_2 = -\boldsymbol{\nu}_1$. 

~\cite{gillespie2000} points out that when two fairly generic dynamical conditions are satisfied, the propensity functions can be used to write down SDEs that describe the temporal evolution of the state variables $\boldsymbol{x}$, and hence a Fokker-Planck equation for $P(\boldsymbol{x},t)$. Here, following~\cite{gillespie2000}, we present the most crucial steps and approximations for this derivation.   

To write the time evolution of the PDF of $\boldsymbol{N}$--- {\it i.e.}, $P(\boldsymbol{N},t\,|\,\boldsymbol{N}_0,t_0)$--- we consider a time scale sufficiently small that the probability of two or more reactions occurring is negligible compared to that of single reaction. We can then simply account for all the the mutually exclusive ways of transitioning in-to or out-of state $\boldsymbol{N}$ via zero or one reaction. The result is that the probability of the system being in state $\boldsymbol{N}$ at time $t+dt$ is given by:
\begin{equation}
P(\boldsymbol{N},t+dt\,|
,\boldsymbol{N}_0,t_0) = P(\boldsymbol{N},t\,\vert\,\boldsymbol{N}_0,t_0) \left[1-\sum_{j=1}^r a_j(\boldsymbol{N})\,dt \right] + \sum_{j=1}^r[P(\boldsymbol{N}-\boldsymbol{\nu}_j,t\,|\,\boldsymbol{N}_0,t_0)\,a_j(\boldsymbol{N}-\boldsymbol{\nu}_j)\,dt].
\label{eq:chemicalMasterProbability}
\end{equation}
The first term in the above equation represents the probability that the system is already in state $\boldsymbol{N}$ and no change occurs; this happens when none of the reactions, $R_j$, take place in the time  interval $\left[t,t+dt\right]$, and corresponds to the probability $1-\sum_{j=1}^r a_j(\boldsymbol{N})dt$. The second term represents the probability that the system is in the state $\boldsymbol{N}$ due to a jump from from one of the states $\boldsymbol{N}-\nu_j$ during the interval $\left[t,t+dt\right]$. 

Rearranging the terms in Eq.~\eqref{eq:chemicalMasterProbability} and taking the limit $dt \to 0$ gives rise to the \textit{chemical master equation}:
\begin{equation}
\frac{\partial}{\partial t}P(\boldsymbol{N},t\,|\,\boldsymbol{N}_0,t_0) = \sum_{j=1}^r[P(\boldsymbol{N}-\boldsymbol{\nu}_j,t\,|\,\boldsymbol{N}_0,t_0)\,a_j(\boldsymbol{N}-\boldsymbol{\nu}_j) - P(\boldsymbol{N},t\,\vert\,\boldsymbol{N}_0,t_0) \,a_j(\boldsymbol{N})]. 
\label{eq:chemicalMasterEquation}
\end{equation}
\noindent In general, this integro-differential equation for the temporal evolution of $P(\boldsymbol{N},t\,|\,\boldsymbol{N}_0,t_0)$ is difficult to solve, which is why we seek a method to instead write down approximate (chemical) Langevin equations.

\textit{Langevin Equations:} The state of the system after a time $\tau$ has elapsed will depend on the number of reactions that have taken place during that time. Let $\mathit{K}_j(\boldsymbol{N},\tau)$ be the number of $R_j$ reactions that occur in the time interval [$t,t+\tau$]. Since each $R_j$ reaction contributes a change $\nu_{ji}$ to the species $X_i$, the number of $X_i$ molecules at time $t+\tau$ will be
\begin{equation}
N_i(t+\tau) = N_i(t) + \sum_{j=1}^r \mathit{K}_j(\boldsymbol{N},\tau)\nu_{ji},\ \textrm{for}\,i = 1,...,n.
\label{eq:changeInSpeciesChemicalLangevin}
\end{equation}
\noindent It is possible to approximate  $\mathit{K}_j(\boldsymbol{N},\tau)$ for any $\tau > 0$ when the following two conditions are satisfied:
\textit{Condition (i)}: The first condition requires $\tau$ to be sufficiently small that none of the propensity functions change considerably, {\it i.e.},
\begin{equation}
a_j(\boldsymbol{N}(t')) \cong a_j(\boldsymbol{N}(t)),\ \boldsymbol{\forall} \:t' \:\in \:[t,t+\tau], \ \boldsymbol{\forall}\: j \:\in \:[1,r]. 
\label{eq:condition1}
\end{equation}
This implies that we choose $\tau$ such that none of the species concentrations change appreciably due to occurrence of any reaction, and that all the reaction events occur independently of each other in the time interval [t,t+$\tau$]. Moreover, the quantities $\mathit{K}_j(\boldsymbol{N},\tau)$, the number of times each reaction occurs in this time interval, must be statistically independent Poisson random variables, denoted by $\mathcal{P}_j(a_j(\boldsymbol{N}),\tau)$. As a result, Eq.~(\ref{eq:changeInSpeciesChemicalLangevin}) can be approximated by
\begin{equation}
N_i(t+\tau) = N_i(t) + \sum_{j=1}^r \mathcal{P}_j(a_j(\boldsymbol{N}),\tau)\nu_{ji}, \ \textrm{for}\, i = 1,...,n.
\label{eq:PoissonApproximation}
\end{equation}

\textit{Condition (ii)}: In addition to the condition of $\tau$ being sufficiently small, we require $\tau$ to be large enough so that 
\begin{equation}
\langle\mathcal{P}_j(a_j(\boldsymbol{N}),\tau)\rangle = a_j(\boldsymbol{N})\tau \gg 1.
\label{eq:condition2}
\end{equation}
As a result, each Poisson random variable, $\mathcal{P}_j(a_j(\boldsymbol{N}),\tau)$, can be approximated by a \textit{normal} random variable with a mean and a standard deviation both equal to $a_j(\boldsymbol{N})\tau$. We denote such a normal random variable by $\mathcal{N}_j(a_j(\boldsymbol{N})\tau,a_j(\boldsymbol{N})\tau)$, where the first argument indicates the mean and the second argument the variance. Therefore, Eq.~\eqref{eq:PoissonApproximation} reduces to
\begin{equation}
N_i(t+\tau) = N_i(t) + \sum_{j=1}^r \mathcal{N}_j(a_j(\boldsymbol{N})\tau,a_j(\boldsymbol{N})\tau)\,\nu_{ji}, \ \textrm{for}\; i = 1,...,n.
\label{eq:NormalApproximation}
\end{equation}
Using the property that $ \mathcal{N}(m,\sigma^2) = m + \sigma\mathcal{N}(0,1)$ and writing $\mathcal{N}(0,1) = \eta(t)$ for any $t$, we have
\begin{equation}
N_i(t+\tau) = N_i(t) + \sum_{j=1}^r \nu_{ji} a_j(\boldsymbol{N})\,\tau  +\sum_{j=1}^r \nu_{ji}[a_j(\boldsymbol{N})\,\tau]^{1/2}\mathcal{\eta}_j(t), \ \textrm{for}\; i = 1,...,n.
\label{eq:linearCombinationNormal}
\end{equation}
Let us denote $\tau$ by dt, and formally consider the infinitesimal limit $dt\to 0$. We also note that $\mathcal{\eta}_j(t)$ and $\mathcal{\eta}_{j'}(t')$ are statistically uncorrelated for $t \neq t'$ and $j \neq j'$, therefore,
\begin{equation}
\frac{d N_i(t)}{dt} = \sum_{j=1}^r \nu_{ji} \,a_j(\boldsymbol{N})  + \sum_{j=1}^r \nu_{ji}\,[a_j(\boldsymbol{N})]^{1/2}\mathcal{\eta}_j(t), \ \textrm{for}\; i = 1,...,n.
\label{eq:chemicalLangevin}
\end{equation}
Dividing the whole equation by $N$ (total number of individuals in the system) and using the property that $a_j(\boldsymbol{N})$ = $Na_j(\boldsymbol{x})$, where $\boldsymbol{x} = \boldsymbol{N}/N$ we finally arrive at the Langevin equations used in the main body of the Chapter:
\begin{equation}
\frac{d x_i}{dt} = \sum_{j=1}^r \nu_{ji} a_j(\boldsymbol{x})  + \frac{1}{\sqrt[]{N}}\sum_{j=1}^r \nu_{ji}[a_j(\boldsymbol{x})]^{1/2}\mathcal{\eta}_j(t), \ \textrm{for}\; i = 1,...,n.
\label{eq:chemicalLangevinConcentration}
\end{equation}

\section{Pairwise interaction model in two spatial dimensions}
\label{app:2d-pairwise}

In this Appendix, we extend the pairwise interaction model to two spatial dimensions. As in the main body of the Chapter, we show how to construct coarse-grained equations for the dynamics of an order parameter from microscopic rules via two methods: ({\it i}) the system-size expansion method and ({\it ii}) the chemical Langevin equation approach. We show that, in contrast to the one-dimensional case, two spatial dimensions are enough to differentiate between the two approaches; the system-size expansion has significant limitations, whereas the CLE approach yields coarse-grained equations with relative ease.

Our starting point is to assume that individuals forage from four sources, located along the positive $x$, positive $y$, negative $x$, and negative $y$ axes. We label these four perpendicular directions $1$ to $4$, respectively. We chose this discretization in lieu of a continuous two-dimensional space for reasons of analytical tractability; in a continuous two-dimensional system there are infinitely many directions along which individuals can move, and thus the number of states to consider are also infinite, prohibiting a straightforward extension of the methods we have discussed for the one-dimensional pairwise interaction model. [For examples of so-called `off-lattice' techniques, see~\citep{othmer1988models,dyson2012macroscopic,dyson2015}].

As before, an ant moving in direction $j$ is denoted by $X_j$.  It changes its direction based on two types of reaction: a copying interaction 
\begin{equation}
X_i + X_j \xrightarrow {c} 2 \, X_{j/i},\ \textrm{for} \; \: i \neq j \; \textrm{and} \; i, j \; \in \; \{1,4 \},
\label{eq:copying-2d}
\end{equation}
where $c$ denotes the specific copying rate; and a spontaneous change in direction
\begin{equation}
X_i \xrightarrow {s} X_j,\ \textrm{for} \; i \neq j \; \textrm{and} \; i, j \; \in \; \{1,4 \},
\label{eq:stochastic-2d}
\end{equation}
where $s$ denotes the specific spontaneous direction switching rate. The proportion of ants in each of the four directions, denoted by $x_i$ for $i \in \{1,2,3,4\}$, satisfy the constraint $\sum_{i=1}^4 x_i = 1$.  We construct equations for the variable $\boldsymbol{x} = \{x_1,x_2,x_3,x_4\}^\mathsf{T}$, representing the proportion of ants moving/foraging in the four different directions $1,...,4$. We then calculate the group polarisation, or order parameter $m = \sqrt{(x_1-x_3)^2 + (x_2 - x_4)^2}$. Each of the state variables $x_i$ can take values in the range $\left[0,1\right]$. The order parameter too, takes values from 0 to 1, with $0$ representing disordered motion and $1$ representing highly aligned motion, or consensus among ants for the foraging source.

\subsection{van-Kampen's system-size expansion of transition rates}

\textit{Transition rates}: Using the shorthand notation $x_i^{+}=x_i + 1/N$ and $x_j^{-}=x_j - 1/N$, the transition rates are the same as for the one dimensional case
\begin{equation}
\label{transitionmij}
T_{ij}(\boldsymbol{x}) = T(x_i^{+},x_j^{-}\,|\,x_i, x_j) = c x_i x_j + s x_j,
\end{equation}
except the indices can now take the values $1,2,3,4$.

\textit{Master Equation}: The corresponding master equation is then just
\begin{equation}
\label{masterEq}
\frac{\partial P(\boldsymbol{x},t)}{\partial t} =   \sum_i \sum_{j\neq i} \left[ P(x_i^+,x_j^-)\:T(x_i,x_j\,|\,x_i^+,x_j^-) -  P(x_i,x_j)\:T(x_i^+,x_j^-\,|\,x_i,x_j) \right].
\end{equation}
As before, this can be simplified by using the step operators $\mathcal{E}^+_i$ and $\mathcal{E}^-_i$, such that
\begin{equation}
\label{introstepOperator}
\mathcal{E}_i^{\pm}(x_i) = f(x_i \pm 1/N) = f(x_i^+),
\end{equation}
whereby the master equation may be re-written to give
\begin{equation}
\frac{\partial P(\boldsymbol{x},t)}{\partial t} =   \sum_i \sum_{j\neq i} \:(\mathcal{E}_i^- \mathcal{E}_j^+ - 1) P(\boldsymbol{x},t)\:T_{ij}.
\label{finalMasterOperator}
\end{equation}

\textit{Fokker-Planck Equation}: Assuming a large-but-finite $N$, we approximate the action of the step-operators via Taylor expansion.  Ignoring terms at $O(1/N^3)$, we have
\begin{subequations}
\begin{equation}
\label{taylorExpand1}
\mathcal{E}_i^+ f(x_i) = f(x_i+\frac{1}{N}) = \left(1+\frac{1}{N}\frac{\partial}{\partial x_i}+\frac{1}{2N^2}\frac{\partial^2}{\partial x_i^2}\right) f(x_i),
\end{equation}
\begin{equation}
\label{taylorExpand2}
\mathcal{E}_i^- f(x_i) = f(x_i-\frac{1}{N}) = \left(1-\frac{1}{N}\frac{\partial}{\partial x_i}+\frac{1}{2N^2}\frac{\partial^2}{\partial x_i^2}\right) f(x_i),
\end{equation}
\end{subequations}
and therefore
\begin{equation}
\label{stepOperatorExpand}
\mathcal{E}_i^- \mathcal{E}_j^+ = 1 + \frac{1}{N}\left(\frac{\partial}{\partial x_j} - \frac{\partial}{\partial x_i}\right) + \frac{1}{2N^2}\left(\frac{\partial}{\partial x_j} - \frac{\partial}{\partial x_i}\right)^2.
\end{equation}
Substituting into \eqref{finalMasterOperator}, rearranging and making the substitution $t = N\tau$, we get the generic Fokker-Planck equation
\begin{equation}
\label{eq:fokkerPlanck}
\frac{\partial P(\boldsymbol{x})}{\partial \tau} =  - \sum_{i = 1}^{n-1} \frac{\partial}{\partial x_i} P(\boldsymbol{x}) A_i(\boldsymbol{x}) + \frac{1}{2} \sum_{i,j = 1}^{n-1} \frac{\partial^2}{\partial x_i \partial x_j} P(\boldsymbol{x}) B_{ij}(\boldsymbol{x}), 
\end{equation}
where $n=4$ is the total number of states/directions/food sources. Due to the constraint $\sum_{i=1}^4 x_i = 1$, we require only three variables to determine the state of the system, and the deterministic vector $\mathcal{A}(\boldsymbol{x})$ therefore has three elements
\begin{equation}
\label{eq:driftCoef}
 \mathcal{A}_i(\boldsymbol{x}) = \sum_{j \neq i}^3 [T_{ij}(\boldsymbol{x}) - T_{ji}(\boldsymbol{x})],\ \mathrm{for}\ i = 1,...,3.
\end{equation}
Similarly, the matrix $\mathcal{B}(\boldsymbol{x})$ is a $3 \times 3$, whose off-diagonal and diagonal elements, respectively, are given by
\begin{eqnarray}
\mathcal{B}_{ij}(\boldsymbol{x}) & = & -\frac{1}{N} [T_{ij}(\boldsymbol{x}) + T_{ji}(\boldsymbol{x})],\ \textnormal{for} \; i \neq j, 
\label{eq:diffusionCoef1} \\
\mathcal{B}_{ii}(\boldsymbol{x}) & = & \frac{1}{N} \sum_{j \neq i} [T_{ij}(\boldsymbol{x}) + T_{ji}(\boldsymbol{x})]. 
\label{eq:diffusionCoef2}
\end{eqnarray}
For the pairwise interaction model in two dimensions represented by \eqref{eq:copying-2d} and \eqref{eq:stochastic-2d}, $\mathcal{A}$ and $\mathcal{B}$ are given by $$\mathcal{A}_i(\boldsymbol{x})= s(1-4x_i),$$ $$\mathcal{B}_{ij}(\boldsymbol{x}) = -\frac{1}{N}[s(x_i+x_j)+2cx_ix_j],$$ and $$\mathcal{B}_{ii}(\boldsymbol{x}) = \frac{1}{N}[s(1+2x_i)+2c\sum_{j \neq i} x_ix_j].$$

\textit{Mesoscopic SDEs}:
Given a multivariate Fokker-Planck equation of the form~\ref{eq:fokkerPlanck}, we can write the corresponding SDEs as follows~\citep{gardiner2009}
\begin{equation}
\label{langevinEq}
\frac{dx_i}{dt} = \mathcal{A}_i(\boldsymbol{x}) + \sum_{j=1}^{n-1} \mathcal{G}_{ij}(\boldsymbol{x})\eta_j(\tau),
\end{equation}
where the $\eta_j$ ($j=1,2,3$) represent uncorrelated Gaussian white noise sources, and the matrix $\mathcal{G}$ is defined by the relation $\mathcal{G}\mathcal{G}^T = \mathcal{B}$. Since all $x_i$ are real, the matrix $\mathcal{G}$ must necessarily be real; the condition for which is that $\mathcal{B}$ is positive definite--- {\it i.e.}, all the eigenvalues of $\mathcal{B}$ are real and positive. For the pairwise interaction model in two dimensions, we find that this condition does not hold. Numerically, we find that one of the eigenvalues of $\mathcal{B}$ is always zero, suggesting that $\mathcal{B}$ is positive {\it semi}-definite, and therefore $\mathcal{G}$ is not unique. We conclude, therefore, that deriving analytically tractable mesoscopic equations is not feasible for the two dimensional pairwise interaction model using the van Kampen's method of system-size expansion. In the next subsection we describe the chemical Langevin equation method, which allows us to write-down a coupled set of SDEs directly from the microscopic reactions.

\subsection{Chemical Langevin approach}

Based on the scheme of reactions in Eqs.~\eqref{eq:copying-2d} and \eqref{eq:stochastic-2d}, we have 12 copying reactions and 12 spontaneous switching reactions. Each of these reactions result in a change in state, such as $x_i\to x_i + 1/N$ and $x_j\to x_j - 1/N$, for various combinations of $i \neq j$. For these 24 different reactions, we can write both propensity functions for the probability of each reaction and a state change vector $\nu_{ji}$, to capture the `stoichiometry' of the reactions. We show all the reactions with their propensity functions and corresponding state change vector entries in Table \ref{tab:chemicalLangevin2D}. Using these details, we show how to write the CLE for the state $x_1$, such that the method can easily be applied to the remaining states. We recall that the generic form of a CLE is given by
\begin{equation}
\frac{dx_i}{dt} = \sum_{j=1}^r\nu_{ji}a_j(\boldsymbol{x}) + \frac{1}{\sqrt[]{N}}\sum_{j=1}^r\nu_{ji}a_j(\boldsymbol{x})^{1/2}\eta_j(t),
\label{eq:chemicalLangevin2D}
\end{equation}
Substituting the values of $\nu_{ji}$ and $a_j(\boldsymbol{x})$ for $i=1$ from Table \ref{tab:chemicalLangevin2D} into the above equation we see that
\begin{table}[h]
\centering
\begin{tabular}{|c|c|c|r|r|r|r|} 
\hline   
S. No. (j)&Reaction&Propensity&$\boldsymbol{\nu}_{j1}$&$\boldsymbol{\nu}_{j2}$&$\boldsymbol{\nu}_{j3}$&$\boldsymbol{\nu}_{j4}$ \\ [0.5ex]   
\hline  
1&$X_1+X_2\xrightarrow{c}2X_1$   &$cx_1x_2$ &1 &-1 &0 &0\\
2&$X_2+X_1\xrightarrow{c}2X_2$   &$cx_1x_2$ &-1 &1 &0 &0\\
3&$X_1+X_3\xrightarrow{c}2X_1$   &$cx_1x_3$ &1 &0 &-1 &0\\
4&$X_3+X_1\xrightarrow{c}2X_3$   &$cx_1x_3$ &-1 &0 &1 &0\\
5&$X_1+X_4\xrightarrow{c}2X_1$   &$cx_1x_4$ &1 &0 &0 &-1\\
6&$X_4+X_1\xrightarrow{c}2X_4$   &$cx_1x_4$ &-1 &0 &0 &1\\
7&$X_2+X_3\xrightarrow{c}2X_2$   &$cx_2x_3$ &0 &1 &-1 &0\\
8&$X_3+X_2\xrightarrow{c}2X_3$   &$cx_2x_3$ &0 &-1 &1 &0\\
9&$X_2+X_4\xrightarrow{c}2X_2$   &$cx_2x_4$ &0 &1 &0 &-1\\
10&$X_4+X_2\xrightarrow{c}2X_4$   &$cx_2x_4$ &0 &-1 &0 &1\\
11&$X_3+X_4\xrightarrow{c}2X_3$   &$cx_3x_4$ &0 &0 &1 &-1\\
12&$X_4+X_3\xrightarrow{c}2X_4$   &$cx_3x_4$ &0 &0 &-1 &1\\
13&$X_1\xrightarrow{s}X_2$   &$sx_1$ &-1 &1 &0 &0\\
14&$X_2\xrightarrow{s}X_1$   &$sx_2$ &1 &-1 &0 &0\\
15&$X_1\xrightarrow{s}X_3$   &$sx_1$ &-1 &0 &1 &0\\
16&$X_3\xrightarrow{s}X_1$   &$sx_3$ &1 &0 &-1 &0\\
17&$X_1\xrightarrow{s}X_4$   &$sx_1$ &-1 &0 &0 &1\\
18&$X_4\xrightarrow{s}X_1$   &$sx_4$ &1 &0 &0 &-1\\
19&$X_2\xrightarrow{s}X_3$   &$sx_2$ &0 &-1 &1 &0\\
20&$X_3\xrightarrow{s}X_2$   &$sx_3$ &0 &1 &-1 &0\\
21&$X_2\xrightarrow{s}X_4$   &$sx_2$ &0 &-1 &0 &1\\
22&$X_4\xrightarrow{s}X_2$   &$sx_4$ &0 &1 &0 &-1\\
23&$X_3\xrightarrow{s}X_4$   &$sx_3$ &0 &0 &-1 &1\\
24&$X_4\xrightarrow{s}X_3$   &$sx_4$ &0 &0 &1 &-1\\
[1ex] 
\hline                          
\end{tabular}
\caption{Reactions in the pairwise interaction model in two dimensions and their propensities and state change vector} 
\label{tab:chemicalLangevin2D}
\end{table}
\begin{equation}
\begin{split}
\sum_{j=1}^r \nu_{j1} a_j(\boldsymbol{x}(t)) 
& = c x_1 x_2 - c x_1 x_2 + c x_1 x_3 - c x_1 x_3 + c x_1 x_4 - c x_1 x_4 - s x_1 + s x_2 - s x_1 + s x_3 - s x_1 + s x_4, \nonumber \\
& = s (x_2 - x_1) + s (x_3 - x_1) + s (x_4 - x_1), \nonumber \\
& = s (1-4x_1).
\end{split}
\end{equation}
And similarly,
\begin{equation}
\begin{split}
\sum_{j=1}^r \nu_{j1} [a_j(\boldsymbol{x}(t))]^{1/2}
& = \sqrt[]{c x_1 x_2}\eta_1 - \sqrt[]{c x_1 x_2}\eta_2 + \sqrt[]{c x_1 x_3}\eta_3 - \sqrt[]{c x_1 x_3}\eta_4 + \sqrt[]{c x_1 x_4}\eta_5 - \sqrt[]{c x_1 x_4}\eta_6 \\
& - \sqrt[]{sx_1}\eta_{13} + \sqrt[]{sx_2}\eta_{14} - \sqrt[]{sx_1}\eta_{15} + \sqrt[]{sx_3}\eta_{16}
- \sqrt[]{sx_1}\eta_{17} + \sqrt[]{sx_4}\eta_{18}, \nonumber
\end{split}
\end{equation}
which results in the following SDE for the variable $x_1$:
\begin{equation}
\begin{split}
\frac{dx_1}{dt} = s (1-4x_1) + \frac{1}{\sqrt[]{N}}(\,& \sqrt[]{c x_1 x_2}\eta_1 - \sqrt[]{c x_1 x_2}\eta_2 + \sqrt[]{c x_1 x_3}\eta_3 - \sqrt[]{c x_1 x_3}\eta_4 + \sqrt[]{c x_1 x_4}\eta_5 - \sqrt[]{c x_1 x_4}\eta_6 \\
& - \sqrt[]{sx_1}\eta_{13} + \sqrt[]{sx_2}\eta_{14} - \sqrt[]{sx_1}\eta_{15} + \sqrt[]{sx_3}\eta_{16}
- \sqrt[]{sx_1}\eta_{17} + \sqrt[]{sx_4}\eta_{18}),
\end{split}
\end{equation}
Performing the similar substitutions for the remaining variables results in
\begin{subequations}
\begin{equation}
\begin{split}
\frac{dx_2}{dt} = s (1-4x_2) + \frac{1}{\sqrt[]{N}}(\,& \sqrt[]{c x_1 x_2}\eta_2 - \sqrt[]{c x_1 x_2}\eta_1 + \sqrt[]{c x_2 x_3}\eta_7 - \sqrt[]{c x_2 x_3}\eta_8 + \sqrt[]{c x_2 x_4}\eta_9 - \sqrt[]{c x_2 x_4}\eta_{10} \\
& + \sqrt[]{sx_1}\eta_{13} - \sqrt[]{sx_2}\eta_{14} - \sqrt[]{sx_2}\eta_{19} + \sqrt[]{sx_3}\eta_{20}
- \sqrt[]{sx_2}\eta_{21} + \sqrt[]{sx_4}\eta_{22}),
\end{split}
\end{equation}
\begin{equation}
\begin{split}
\frac{dx_3}{dt} = s (1-4x_3) + \frac{1}{\sqrt[]{N}}(\,& \sqrt[]{c x_1 x_3}\eta_4 - \sqrt[]{c x_1 x_3}\eta_3 + \sqrt[]{c x_2 x_3}\eta_8 - \sqrt[]{c x_2 x_3}\eta_7 + \sqrt[]{c x_3 x_4}\eta_{11} - \sqrt[]{c x_3 x_4}\eta_{12} \\
& + \sqrt[]{sx_1}\eta_{15} - \sqrt[]{sx_3}\eta_{16} + \sqrt[]{sx_2}\eta_{19} - \sqrt[]{sx_3}\eta_{20}
- \sqrt[]{sx_3}\eta_{23} + \sqrt[]{sx_4}\eta_{24}),
\end{split}
\end{equation}
\begin{equation}
\begin{split}
\frac{dx_4}{dt} = s (1-4x_4) + \frac{1}{\sqrt[]{N}}(\,& \sqrt[]{c x_1 x_4}\eta_6 - \sqrt[]{c x_1 x_4}\eta_5 + \sqrt[]{c x_2 x_4}\eta_{10} - \sqrt[]{c x_2 x_4}\eta_9 + \sqrt[]{c x_3 x_4}\eta_{12} - \sqrt[]{c x_3 x_4}\eta_{11} \\
& + \sqrt[]{sx_1}\eta_{17} - \sqrt[]{sx_4}\eta_{18} + \sqrt[]{sx_2}\eta_{21} - \sqrt[]{sx_4}\eta_{22}
+ \sqrt[]{sx_3}\eta_{23} - \sqrt[]{sx_4}\eta_{24}),
\end{split}
\end{equation}
\label{eq:langevin-2d}
\end{subequations}
where, $\eta_1,\eta_2,\ldots,\eta_{24}$ are delta-correlated Gaussian white noise sources. Unfortunately, further simplification of the stochastic terms is not possible in this case because the coupled nature of the equations.  Nevertheless, the corresponding multivariate Fokker-Planck equation can be derived by following~\cite{gillespie2002chemical}.
\begin{equation}
\begin{split}
\frac{\partial P(\boldsymbol{x},t|\boldsymbol{x}_0,t_0)}{\partial t}
& = -\sum_{i=1}^n\frac{\partial}{\partial x_i}\Bigg[\left(\sum_{j=1}^r\nu_{ji}a_j(\boldsymbol{x})\right)P(\boldsymbol{x},t|\boldsymbol{x}_0,t_0)\Bigg] \\
& + \frac{1}{2N} \sum_{i=1}^n\frac{\partial^2}{\partial x_i^2}\Bigg[\left(\sum_{j=1}^r\nu_{ji}^2a_j(\boldsymbol{x})\right)P(\boldsymbol{x},t|\boldsymbol{x}_0,t_0)\Bigg] \\
& + \frac{1}{N} \sum_{i'=1}^n\sum_{i<i'}\frac{\partial^2}{\partial x_i\partial x_{i'}} \Bigg[\left(\sum_{j=1}^r\nu_{ji}\nu_{ji'}a_j(\boldsymbol{x})\right)P(\boldsymbol{x},t|\boldsymbol{x}_0,t_0)\Bigg].
\end{split}
\label{eq:langevin-to-FP}
\end{equation}
Substituting the values of $\nu_{ji}$ and $a_j(\boldsymbol{x})$ from Table \ref{tab:chemicalLangevin2D} recovers a Fokker-Planck equation, of the form
\begin{equation}
\frac{\partial P(\boldsymbol{x})}{\partial \tau} =  - \sum_{i = 1}^{n} \frac{\partial}{\partial x_i} \left[P(\boldsymbol{x}) A_i(\boldsymbol{x})\right] + \frac{1}{2} \sum_{i,j = 1}^{n} \frac{\partial^2}{\partial x_i \partial x_j} \left[P(\boldsymbol{x}) B_{ij}(\boldsymbol{x})\right],
\label{eq:Fokker-Planck-chemical-Langevin-2d}
\end{equation}
where $\mathcal{A}(\boldsymbol{x})$ is given by
$$\mathcal{A}_i(\boldsymbol{x})= s(1-4x_i),$$ whilst the diagonal and off-diagonal elements of $\mathcal{B}(\textbf{x})$ are given by $$\mathcal{B}_{ii}(\boldsymbol{x}) = \frac{1}{N}[s(1+2x_i)+2c\sum_{j \neq i} x_ix_j],$$ and $$\mathcal{B}_{ij}(\boldsymbol{x}) = -\frac{1}{N}[s(x_i+x_j)+2cx_ix_j],$$ respectively. This is the same Fokker-Planck equation as derived using system-size expansion.

\nomenclature{$X_i$}{An individual in a given state/direction/food source $i$}
\nomenclature{$N_i$}{Number of individuals in $i_{th}$ state/direction/food source}
\nomenclature{$x_i$}{Proportion of individuals in $i_{th}$ state/direction/food source}
\nomenclature{$N$}{Total number of individuals in the system or population, also referred as {\it system-size}}
\nomenclature{$n$}{Total number of states/directions/food sources in the system}
\nomenclature{$r$}{Total number of reactions/rules according to which individuals can change their state from $X_i$ to $X_j$}
\nomenclature{$s$}{Rate of reaction of spontaneous switching between states/directions/food sources}
\nomenclature{$c$}{Rate of reaction pertaining to copying/feedback interaction amongst individuals}
\nomenclature{$h$}{Rate of reaction pertaining to higher-order interaction amongst individuals}
\nomenclature{$\boldsymbol{N}$}{State of the system, consisting of number of individuals $(N_i)$ in different states/directions/food sources $(\boldsymbol{N} \equiv$ $\{N_1,...,N_n\}^\mathsf{T})$}
\nomenclature{$\boldsymbol{x}$}{State of the system, consisting of proportions of individuals $(x_i)$ in different states/directions/food sources $(\boldsymbol{x} \equiv$ $\{x_1,...x_n\}^\mathsf{T})$}
\nomenclature{$P_s$}{Steady state probability density function}
\nomenclature{$m$}{Order parameter representing by the amount of consensus amongst individuals to choose a direction or food source}
\nomenclature{$\eta$}{Uncorrelated Gaussian white noise process with mean zero and variance of one}
\clearpage
\section*{References}

\bibliographystyle{elsarticle-harv}

\end{document}